\documentclass[iop,tighten]{emulateapj}
\usepackage{apjfonts}
\usepackage{multirow}
\usepackage{xspace}
\usepackage{amsmath}
\usepackage{amssymb}
\usepackage{amsthm}
\usepackage{amscd}
\usepackage{amsfonts}
\usepackage{balance}
\usepackage{hyperref} 
\usepackage{nag}
\usepackage{color}
\usepackage[usenames,dvipsnames]{xcolor}

\def\Msun{M_\odot}
\def\kb{k_{\mathrm{B}}}

\def\keV{\mathrm{keV}}
\def\Mpc{\mathrm{Mpc}}
\def\kpc{\mathrm{kpc}}
\def\kms{\mathrm{km\,s^{-1}}}

\def\gas{\mathrm{gas}}
\def\vir{\mathrm{vir}}

\def\ta{\mathrm{ta}}

\hypersetup{%
    colorlinks=false, 
    urlcolor=blue, 
    linkcolor=blue, 
    citecolor=blue,
    breaklinks=true, 
    bookmarksnumbered=true,
    bookmarksopen=false, 
    bookmarksopenlevel=0,
    hypertexnames=true, 
    unicode=true,          
    pdftoolbar=true,        
    pdfmenubar=true,        
    pdffitwindow=false,     
    pdfstartview={FitH}    
}

\usepackage{etoolbox}
\makeatletter

\patchcmd{\NAT@citex}
  {\@citea\NAT@hyper@{\NAT@nmfmt{\NAT@nm}\NAT@date}}
  {\@citea\NAT@nmfmt{\NAT@nm}\NAT@hyper@{\NAT@date}}
  {}
  {}

\patchcmd{\NAT@citex}
  {\@citea\NAT@hyper@{%
     \NAT@nmfmt{\NAT@nm}%
     \hyper@natlinkbreak{\NAT@aysep\NAT@spacechar}{\@citeb\@extra@b@citeb}%
     \NAT@date}}
  {\@citea\NAT@nmfmt{\NAT@nm}%
   \NAT@aysep\NAT@spacechar%
   \NAT@hyper@{\NAT@date}}
  {}
  {}

\patchcmd{\NAT@citex}
  {\@citea\NAT@hyper@{%
     \NAT@nmfmt{\NAT@nm}%
     \hyper@natlinkbreak{\NAT@spacechar\NAT@@open\if*#1*\else#1\NAT@spacechar\fi}%
       {\@citeb\@extra@b@citeb}%
     \NAT@date}}
  {\@citea\NAT@nmfmt{\NAT@nm}%
   \NAT@spacechar\NAT@@open\if*#1*\else#1\NAT@spacechar\fi%
   \NAT@hyper@{\NAT@date}}
  {}
  {}

\makeatother

\begin{document}
\title{Mass Accretion and its Effects on the Self-Similarity of Gas Profiles \\ in the Outskirts of Galaxy Clusters}
\shorttitle{Gas Profiles in Cluster Outskirts}
\shortauthors{Lau et al}
\journalinfo{The Astrophysical Journal, 806:68 (12pp), 2015 June 10} 
\submitted{Received 2014 November 19; accepted 2015 April 21; published 2015 June 9}

\author{Erwin T. Lau\altaffilmark{1,2}}
\author{Daisuke Nagai\altaffilmark{1,2,3}}
\author{Camille Avestruz\altaffilmark{1,2}}
\author{Kaylea Nelson\altaffilmark{2,3}}
\author{Alexey Vikhlinin\altaffilmark{4}}

\affil{ 
\textsuperscript{1}Department of Physics, Yale University, New Haven, CT 06520, USA; \href{mailto:erwin.lau@yale.edu}{erwin.lau@yale.edu} \\
\textsuperscript{2}Yale Center for Astronomy and Astrophysics, Yale University, New Haven, CT 06520, USA \\
\textsuperscript{3}Department of Astronomy, Yale University, New Haven, CT 06520, USA \\
\textsuperscript{4}Harvard-Smithsonian Center for Astrophysics, 60 Garden Street, Cambridge, MA 02138, USA
}


\begin{abstract}
Galaxy clusters exhibit remarkable self-similar behavior which allows us to establish simple scaling relationships between observable quantities and
cluster masses, making galaxy clusters useful cosmological probes. Recent X-ray observations suggested that self-similarity may be broken in the outskirts of galaxy clusters. In this work, we analyze a mass-limited sample of massive galaxy clusters from the {\em Omega500} cosmological hydrodynamic simulation to investigate the self-similarity of the diffuse X-ray emitting intracluster medium (ICM) in the outskirts of galaxy clusters.  We find that the self-similarity of the outer ICM profiles is better preserved if they are normalized with respect to the mean density of the universe, while the inner profiles are more self-similar when normalized using the critical density.  However, the outer ICM profiles as well as the location of accretion shock around clusters are sensitive to their mass accretion rate, which causes the apparent breaking of self-similarity in cluster outskirts.  We also find that the collisional gas does not follow the distribution of collisionless dark matter (DM) perfectly in the infall regions of galaxy clusters, leading to $10\%$ departures in the gas-to-DM density ratio from the cosmic mean value. Our results have a number implications for interpreting observations of galaxy clusters in X-ray and through the Sunyaev--Zel'dovich effect, and their applications to cosmology.
\end{abstract}

\section{Introduction}
\label{sec:intro}

Galaxy clusters are the most massive gravitationally bound objects in
the universe.  Most of the baryons are in the form of hot X-ray
emitting gas and reside in the deep gravitational potential wells of
galaxy clusters.  The hot gas is detectable in both the X-ray and the
microwave, through the Sunyaev--Zel'dovich (SZ) effect.
Observations of the intracluster medium (ICM) 
show remarkable regularity over a wide range of mass and redshift, 
making galaxy clusters powerful probes for cosmology 
\citep[e.g.][for review]{allen_etal2011}.  

Self-similarity is a generic prediction of gravitational structure formation.
It features simple scaling relations between the observable properties and 
mass of galaxy clusters \citep[][for a recent review]{kaiser1986,voit2005,kravtsov_borgani2012}.  
When scaled by cluster mass, the radial profiles of thermodynamic
properties of the ICM display remarkable resemblance
\citep{vikhlinin_etal2006, nagai_etal2007a, pratt_etal2009,
  arnaud_etal2010} outside of cluster cores where the effects of
non-gravitational physics are small.  When integrated, the
self-similar ICM quantities can serve as robust observational proxies
for cluster mass, such as its thermal energy content, 
$Y_X$ \citep{kravtsov_etal2006}, or its X-ray luminosity, 
$L_X$ \citep{maughan_etal2007}.  
The outer regions of clusters are ideal for
robust ICM measurements and inferences of cluster mass since these
measurements at large radii are less susceptible to complex
astrophysical processes that affect measurements in the cluster core,
such as gas cooling, star formation, and energy injections from
supernovae and active galactic nuclei.

Both X-ray and SZ observations have recently measured
properties of the ICM out to the virial radius \citep[see, e.g.,][for
review]{reiprich_etal2013}.  However, some of these observations revealed
several unexpected results that deviate from theoretical predictions.
First, hydrodynamical simulations predict that the entropy of the ICM
should have a power law scaling with cluster-centric radius, as the
entropy is set by the accretion shock of the cluster
\citep{tozzi_norman2001,voit_etal2003}.  
But, {\em Suzaku} X-ray measurements of the outer regions of clusters
showed lower and flatter entropy profiles than 
the theoretically predicted power law, breaking the self-similar scaling
\citep[e.g.,][]{george_etal2009,bautz_etal2009,reiprich_etal2009,
hoshino_etal2010,kawaharada_etal2010, akamatsu_etal2011,
walker_etal2013, urban_etal2014,okabe_etal2014}. 
Additionally, {\em Suzaku} measured an enclosed gas
mass fraction of the Perseus cluster that curiously exceeds the
cosmic baryon fraction at large radii \citep{simionescu_etal2011}.
In addition, {\em Chandra} follow-up observations of 
SZ-selected clusters from the South Pole Telescope
survey also found signs of entropy flattening and redshift evolution of the pressure
profile in the outer regions of clusters, indicating signatures of departures
from self-similarity \citep[][but see \citealt{eckert_etal2013a, 
morandi_etal2015}]{mcdonald_etal2014}. 
 
In the prevalent $\Lambda$CDM picture of structure formation, 
galaxy clusters are dynamically young objects that are still growing
via mergers and accretion. The outskirts of galaxy clusters can be
regarded as ``cosmic melting pots,'' where infalling materials are being virialized. 
The accreting gas dissipates heat through shocks and turbulence, establishing 
the overall thermodynamical structures in galaxy clusters.
An improved understanding of the ``cosmic melting pot'' will advance
our ability to use clusters for precision cosmology, particularly in
light of ongoing and upcoming multi-wavelength cluster surveys,
including {\em Planck} and {\em eROSITA}.

A variety of astrophysical processes in cluster outskirts can contribute to the observed 
deviations from self-similar predictions.  Previous theoretical works indicated that processes 
such as gas inhomogeneities
\citep{nagai_lau2011,zhuravleva_etal2013,vazza_etal2013,roncarelli_etal2013,
  rasia_etal2014} and non-thermal pressure support
\citep[e.g,][]{nelson_etal2014b, shi_komatsu2014, shi_etal2014}
may play significant roles.  These non-equilibrium 
processes are likely to be driven by mass accretion in the outskirts of clusters. 

Recent $N$-body simulations suggested that dark matter (DM) halos
accrete mass at different rates depending on their mass, redshift
\citep[e.g.,][]{cuesta_etal2008,mcbride_etal2009}.  
Variations in the mass accretion rate (MAR) of halos introduce differences in the DM
density profiles in halo outskirts, leading to apparent deviations
from self-similarity \citep{diemer_kravtsov2014}. 
If gas traces DM, we would expect mass accretion to similarly 
affect the gas distribution in cluster outskirts. 
Understanding the effects of MAR on gas properties in cluster outskirts requires 
cosmological hydrodynamic simulations that self-consistently follow
the dynamics of DM and gas.

In this work, we analyze a mass-limited sample of galaxy clusters extracted from 
the {\em Omega500} cosmological hydrodynamic simulation \citep{nelson_etal2014}. 
We find that the ICM profiles in cluster outskirts are affected by the mass accretion in two ways.
First, the outer ICM profiles are more self-similar with redshift when they are normalized using 
the mean mass density of the universe,  because the mean density 
traces the redshift evolution of MAR better than the critical density.  
Second, at any given redshift, 
the outer ICM profiles and the location of accretion shock around clusters are sensitive to the MAR.   
We also find the collisional gas does not trace accretion of the collisionless DM perfectly, 
leading to departures in the density ratio between gas and DM from the cosmic mean value. 
We discuss implications of our results on X-ray and SZ observations of the ICM profiles in cluster outskirts, 
mass-observable scaling relations, and the use of galaxy clusters as cosmological probes. 

The paper is organized as follows. In Section~\ref{sec:theory} we describe the concepts of cluster 
mass definitions, self-similar model, and MAR. In Section~\ref{sec:sim} we describe our simulated 
cluster sample. In Section~\ref{sec:results} we examine the dependence of the gas profiles on the 
definition of cluster mass and their MAR. We provide our discussion and conclusions in Section~\ref{sec:conclusions}.

\section{Theoretical Considerations}
\label{sec:theory}

\subsection{Cluster Mass Definitions}
\label{sec:masses}

Galaxy cluster forms at the intersections of large-scale filamentary
structures in the universe and they do not have well-defined physical
edge. To calculate cluster masses, the common approach is to define
the boundary of a cluster as a sphere enclosing an average matter
density equal to a reference overdensity $\Delta$ times a reference
background density, $\rho_{\rm ref}$. The mass of the cluster is then
given as,
\begin{equation}
M_{\Delta} \equiv \frac{4\pi}{3}\Delta \rho_{\rm ref}(z) R_{\Delta}^3
\end{equation}
where $R_{\Delta}$ is the radius within which we compute the enclosed mass. 
Two common choices of the background density $\rho_{\rm ref}$ are the critical density, 
$\rho_c(z) \equiv 3H_0^2E^2(z)/(8\pi G)$, and the mean matter density, 
$\rho_m(z) = \rho_c(z) \Omega_m(z)$, in the standard $\Lambda$CDM 
spatially flat cosmological model, 
where $E^2(z) \equiv \Omega_m(1+z)^3+\Omega_{\Lambda}$, 
$\Omega_m(z)=\Omega_m (1+z)^3/E^2(z)$, and $\Omega_m$ 
(without the explicit $z$-dependence) 
refers to the present-day mass density fraction of the universe. 
The reference overdensity $\Delta$ is usually chosen to be 
a value close to $18\pi^2 \approx 178$, 
which corresponds to the virial overdensity in the flat matter dominated, 
Einstein-de Sitter universe ($\Omega_m = 1- \Omega_\Lambda= 1$). 
In the more realistic flat $\Lambda$CDM model, 
the virial overdensity varies with redshift \citep[e.g.,][]{bryan_norman1998}. 

In the literature, $\rho_c(z)$ has been widely used to define cluster masses 
($\rho_{\rm ref} = \rho_c(z)$, with $\Delta = \Delta_c$). 
In particular, $\Delta_c = 500$ has been used in the analyses of {\em Chandra} 
and {\em XMM-Newton} X-ray observations of galaxy clusters, 
since it corresponds to the radius out to which gas density and temperature profiles 
of the ICM can be reliably measured. 
$\Delta_c = 200$ is also adopted in recent \emph{Suzaku} X-ray 
and \emph{Planck} SZ observations which extended the measurements 
of the ICM profiles to larger cluster-centric radii. 
On the other hand, $\rho_m(z) \propto (1+z)^3$ is independent
of other cosmological parameters, e.g., the Hubble parameter. 
Using $\rho_m(z)$ to define DM halos 
also leads to a more ``universal'' mass function \citep[e.g.,][]{white_2002} 
and has been used in calibrating halo mass functions in $N$-body simulations 
\citep[e.g.,][]{jenkins_etal2001,tinker_etal2008}. 

\subsection{Self-similarity}
\label{sec:self-similar-collapse}

In the current hierarchical structure formation model, galaxy cluster of mass $M$ at redshift $z$ 
forms from gravitational collapse of the primordial cosmological density perturbation, 
when its linear density fluctuation $\delta(M,z)$ reaches the collapse threshold $\delta_c=1.686$.
Since the primordial density perturbations are well-characterized by the Gaussian distribution, properties of 
galaxy clusters are uniquely characterized by its density peak height, $\nu(M,z) \equiv \delta_c/\sigma(M,z)$, 
where $\sigma(M,z)$ is the characteristic linear density fluctuation smoothed over mass scale $M$ at redshift $z$. 

Strictly speaking, self-similarity only holds when the initial linear
density fluctuations and their subsequent gravitational collapse into
cluster halos are scale-free, and there is no physical scale
associated with non-gravitational processes operating during cluster
formation. These conditions are approximately true for cluster-size
halos, where linear density fluctuations follow a power law. 
During the subsequent collapse until $z\gtrsim0.5$, 
the effects of dark energy are also assumed to be small, 
keeping cluster growth almost scale-free. Baryonic physics, 
such as radiative cooling, star formation and feedback, 
break self-similarity, but their effects are mostly confined to 
within the cluster core. 

On large scales, the majority of the baryonic component is in the form
of X-ray emitting ICM and is expected to follow the distribution 
of the gravitationally dominant DM. The self-similar model predicts that 
cluster gas profiles for a given mass (or peak height) appear ``universal'' 
when they are scaled with respect to the reference background density 
of the universe \citep[see, e.g.,][]{voit2005}. For example, 
the gas density is scaled using the mean cosmic baryon overdensity, 
defined as $\rho_{\gas,\Delta} \equiv f_b \Delta \rho_{\rm ref}(z)$, 
where $\Delta$ is the redshift independent chosen overdensity, 
$\rho_{\rm ref}(z)$ is the reference mass density of the universe at redshift $z$, 
and $f_b \equiv \Omega_b/\Omega_m$ is the cosmic baryon fraction. 
Similarly, other quantities, such as temperature, pressure, entropy, and velocity, 
can be normalized with appropriate scaling that depends on mass and redshift: 
$\kb T_{\Delta} \equiv GM_{\Delta}\mu m_p/(2R_{\Delta})$,  
$P_{\Delta} \equiv \rho_{\gas,\Delta}\kb T_{\Delta}/(\mu m_p)$,  
$K_{\Delta} \equiv \kb T_{\Delta}/(\mu m_p \rho_{\gas,\Delta}^{2/3})$, 
and $V_{{\rm circ}, \Delta} \equiv \sqrt{GM_{\Delta}/R_{\Delta}}$,
where $G$ is the gravitational constant, 
$m_p$ is the proton mass and $\kb$ is the Boltzmann constant.  
For the cases where the reference background density is set to the critical or the mean density, 
we set $\Delta = \Delta_c$ or $\Delta = \Delta_m \equiv \Delta_c/\Omega_m(z)$, respectively.  
In this paper, we consider two cases: $\Delta_c = 200$ and $\Delta_m = 200$. The 
 the self-similar values of the thermodynamical quantities for $\Delta_c = 200$ are:
 \begin{align}
 &\rho_{200c} = 4.759\times 10^{-14}\,E(z)^{2}\;\Msun\,{\Mpc}^{-3},\\
 &\kb T_{200c} = 8.145\;\keV\, \left(\frac{M_{200c}}{10^{15} h^{-1}M_\odot}\right)^{2/3}E(z)^{2/3},  \\ 
 &P_{200c}  = 2.660\times 10^{-3}\;{\rm keV\; cm^{-3}}\,\left(\frac{M_{200c}}{10^{15} h^{-1}M_\odot}\right)^{2/3}E(z)^{8/3}, \\
 &K_{200c} = 1718\;{\rm keV\,cm^{2}} \left(\frac{M_{200c}}{10^{15} h^{-1}M_\odot}\right)^{2/3} E(z)^{-2/3};
 \end{align}
and for $\Delta_m = 200$:
 \begin{align}
&\rho_{200m} =  1.285\times 10^{-12}\,(1+z)^3\;\Msun\,{\Mpc}^{-3},\\
&\kb T_{200m}  =  5.265\;{\keV} \left(\frac{M_{200m}}{10^{15} h^{-1}M_\odot}\right)^{2/3} (1+z), \\ 
&P_{200m}  = 4.641\times 10^{-4}\;{\rm keV\; cm^{-3}}  \left(\frac{M_{200m}}{10^{15} h^{-1}M_\odot}\right)^{2/3}(1+z)^4,\\
&K_{200m} = 2800\;{\rm keV\; cm^{2}}\, \left(\frac{M_{200m}}{10^{15} h^{-1}M_\odot}\right)^{2/3} (1+z)^{-1}.
\end{align}
We adopt $f_b = 0.1737$ from the {\em WMAP5} cosmology used in this work, and
$\mu=0.59$ to be the mean particle weight of the fully ionized ICM.

\begin{figure*}[t]
\begin{center}
\includegraphics[scale=0.18]{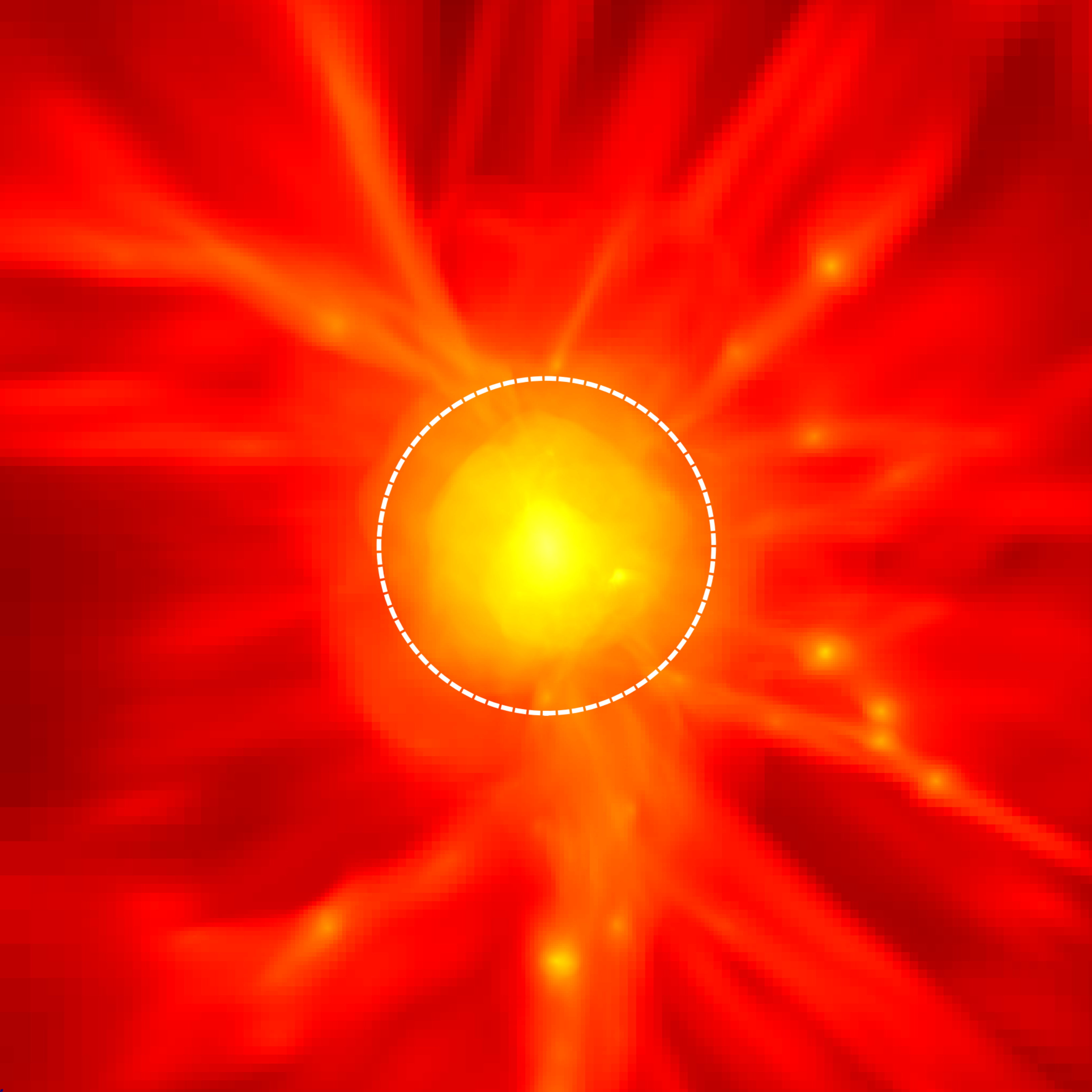}
\includegraphics[scale=0.18]{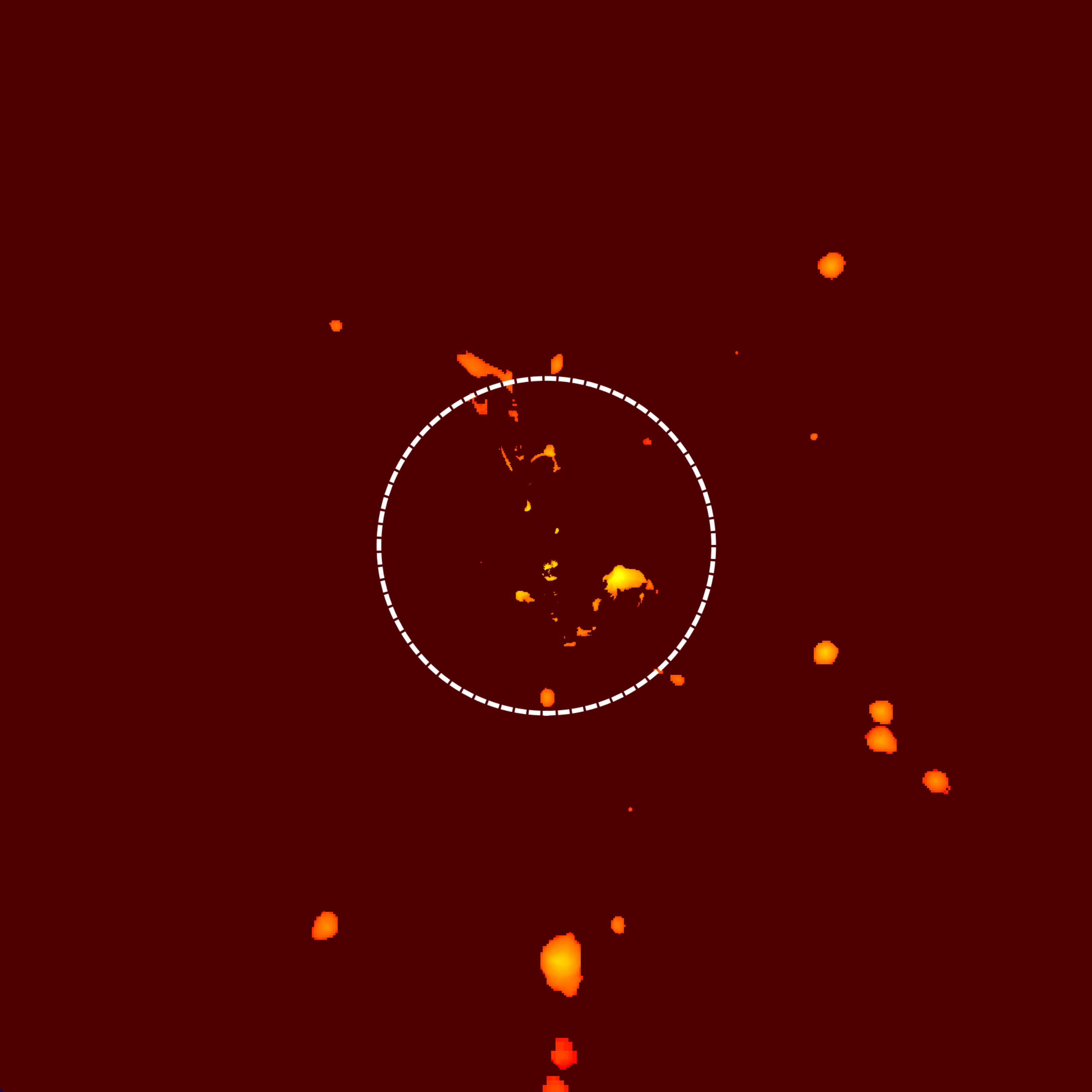}
\includegraphics[scale=0.18]{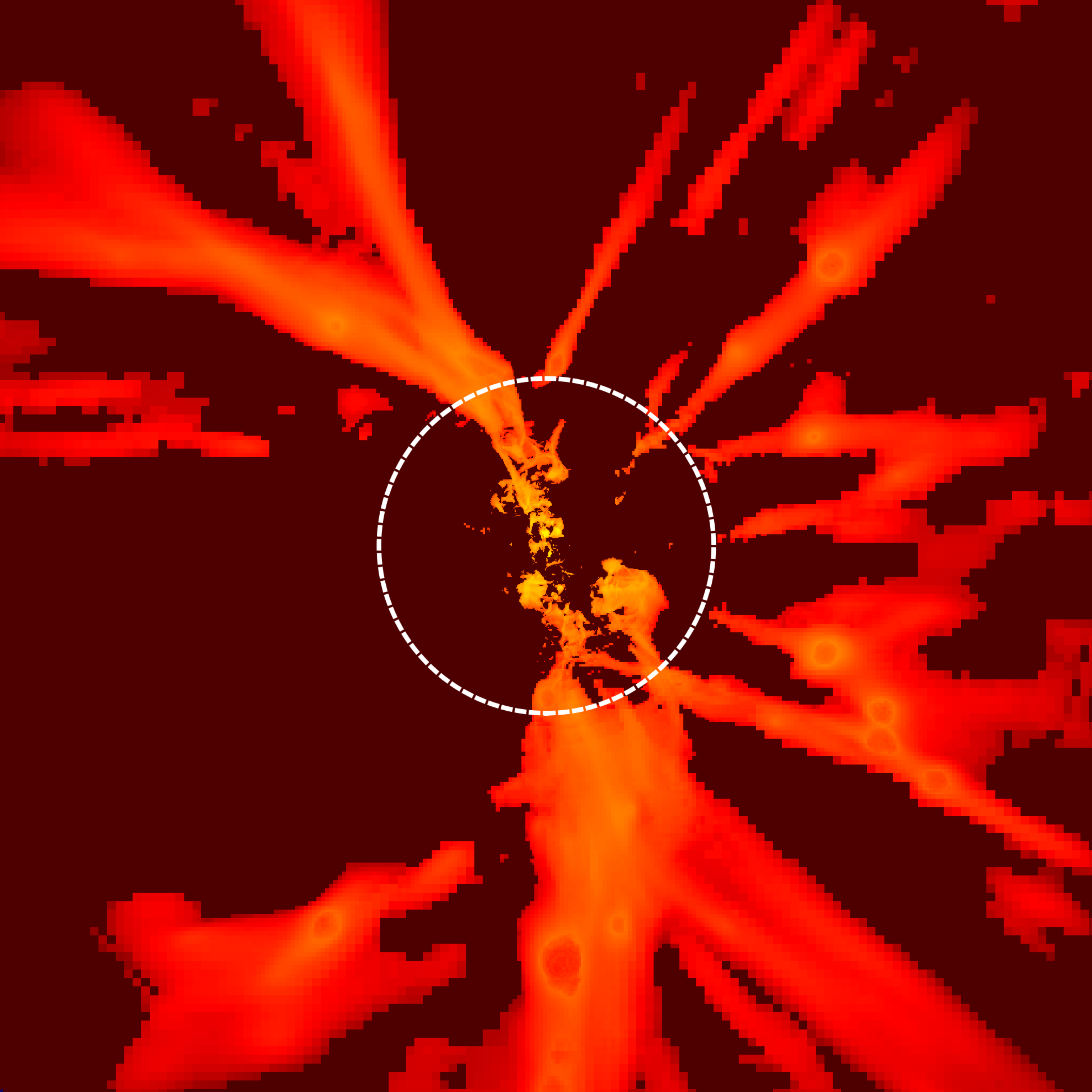}
\caption
{Projected gas density maps of one of the $z=0$ clusters selected from the sample. From {\em left} to {\em right}: map of the total gas, clumps, and filaments. The dimension for each panel is $15.6\,h^{-1}\Mpc\times15.6\,h^{-1}\Mpc$, with depth of $1.9\,h^{-1}\Mpc$. The circle in dashed line shows $R_{200m} = 3.3\,h^{-1}\Mpc$ of the cluster. }
\label{fig:cluster_map}
\end{center}
\end{figure*}

\subsection{Mass Accretion Rate (MAR)}
\label{sec:mar}

In this work, we propose a proxy for the {\em instantaneous} MAR $\alpha_\Delta$, 
defined as the ratio of radial infall velocity to the circular velocity of the cluster halo 
measured at some chosen radius $R_{\alpha}$, 
\begin{equation}\label{eq:alpha}
\alpha_{\Delta} = \frac{V_r^{\rm DM}(r= R_{\alpha}) }{ V_{\rm circ, \Delta} }, 
\end{equation}
where $R_{\alpha}$ is a free parameter denoting a certain radius located inside the infall region where $V_r^{\rm DM}<0$. 
Physically, a halo that is actively accreting in its outskirts will have a large negative value of $\alpha_{\Delta}$.  

\citet{diemer_kravtsov2014} proposed a different MAR proxy $\Gamma_{\Delta}$:
\begin{equation}\label{eq:gamma}
\Gamma_{\Delta} = \frac{\log_{10}\left(M_{\Delta}(a_0)/M_{\Delta}(a_1)\right)}{\log_{10}(a_0/a_1)}
\end{equation}
where $M_{\Delta}(a_0)$ and $M_{\Delta}(a_1)$ are the mass of the halo 
at $a_0=1\,(z=0)$ and its progenitor at $a_1=0.67\,(z=0.5)$ respectively.  
A higher $\Gamma_{\Delta}$ means that the halo has experienced 
a greater {\em physical} mass accretion between the two redshifts. 

Our proposed MAR proxy $\alpha_{\Delta}$ have an advantage over $\Gamma_{\Delta}$ 
because $\alpha_{\Delta}$ can be measured for halos at any given redshift, 
while $\Gamma_{\Delta}$ is an integrated quantity defined between two chosen redshifts. 
$\alpha_{\Delta}$ is also in principle an observable quantity that can be measured from
radial velocities of infalling galaxies or gas in cluster outskirts. 
We compare these two MAR proxies in Section~\ref{sec:mar_dependence}. 

\section{Cosmological Hydrodynamic Simulations}
\label{sec:sim}

\subsection{Data and Halo Selection}

In this work we analyze a sample of simulated galaxy clusters from the 
{\em Omega500} Simulation Project \citep{nelson_etal2014}, 
which is a high-resolution hydrodynamical simulation of a large cosmological volume 
with the comoving box length of $500\,h^{-1}\,\Mpc$. 
The simulation is performed using the Adaptive Refinement Tree (ART) 
$N$-body+gas-dynamics code \citep{kravtsov1999, kravtsov_etal2002, rudd_etal2008}, 
which is an Eulerian code that uses adaptive refinement in space and time, 
and non-adaptive refinement in mass \citep{klypin_etal2001} 
to achieve the dynamic ranges to resolve the cores of halos 
formed in self-consistent cosmological simulations in a flat $\Lambda$CDM model
 with WMAP 5 years ({\em WMAP5}) cosmological parameters: 
$\Omega_m = 1 - \Omega_{\Lambda} = 0.27$, $\Omega_b = 0.0469$, 
$h = 0.7$ and $\sigma_8 = 0.82$, where the Hubble constant is defined as 
$100\,h\;\kms\,\Mpc^{-1}$ and $\sigma_8$ is the mass variance within 
spheres of radius $8\,h^{-1}\Mpc$. The simulation is performed on a 
uniform $512^3$ grid with 8 levels of mesh refinement, implying a 
maximum comoving spatial resolution of $3.8\,h^{-1} \kpc$.  
Our simulations are based on simple non-radiative hydrodynamics, 
allowing us to isolate the effects of MAR on self-similarity 
from the effects of complicated galaxy formation physics, 
which are expected to be small in the outskirts of clusters. 
The current work also serves as a baseline for future studies of the effects of such physics. 

Galaxy clusters are identified in the simulation using a spherical overdensity halo finder 
described in \cite{nelson_etal2014}.  We select clusters with 
$M_{500c} \geq 3\times10^{14}\,h^{-1}\Msun$ at $z=0$ 
and re-simulate the box with higher resolution DM particles in regions of the selected clusters, 
resulting in an effective mass resolution of $2048^{3}$, 
which corresponds to a mass resolution of $1.09 \times 10^9\, h^{-1}\Msun$. 
To study the redshift evolution for the ICM profiles, we extract halos from 
four redshift outputs: $z=0.0, 0.5,1.0,1.5$. At each redshift 
we apply a mass cut to ensure a mass-limited sample by comparing 
our mass function to that of \citet{tinker_etal2008} and setting the mass cut 
to ensure that the sample is complete above the chosen mass threshold. 
The mass-cuts and resulting sample sizes are as follows: 
65 clusters with $M_{200m} \geq 6\times 10^{14} h^{-1}\Msun$ at $z = 0$,  
48 clusters with $M_{200m} \geq 2.5\times 10^{14} h^{-1}\Msun$ at $z = 0.5$,  
42 clusters with $M_{200m} \geq 1.3\times 10^{14} h^{-1}\Msun$ at $z = 1.0$,  
and 42 clusters with $M_{200m} \geq 7 \times 10^{13} h^{-1}\Msun$ at $z = 1.5$.

\begin{figure*}
\begin{center}
\includegraphics[scale=0.65]{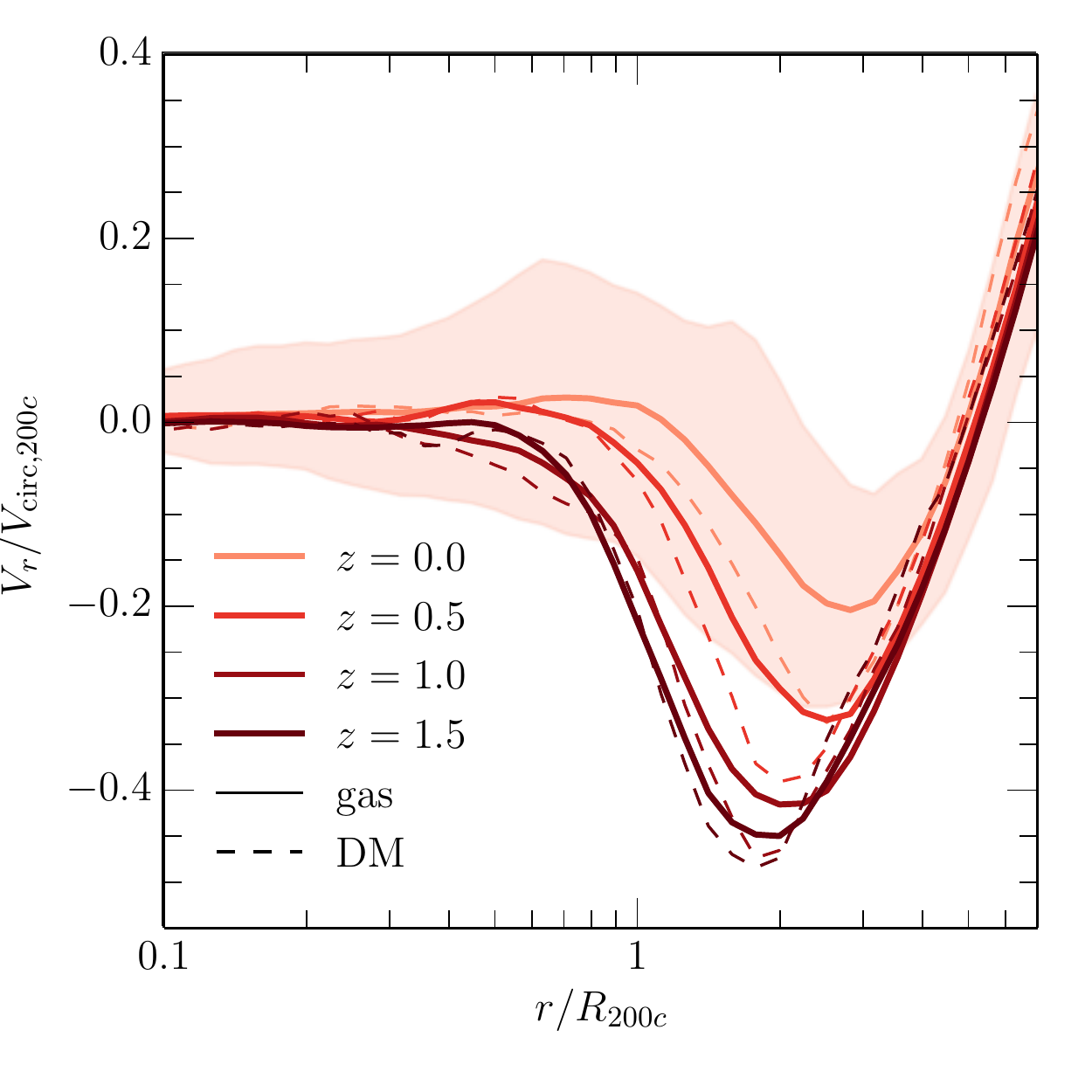}
\includegraphics[scale=0.65]{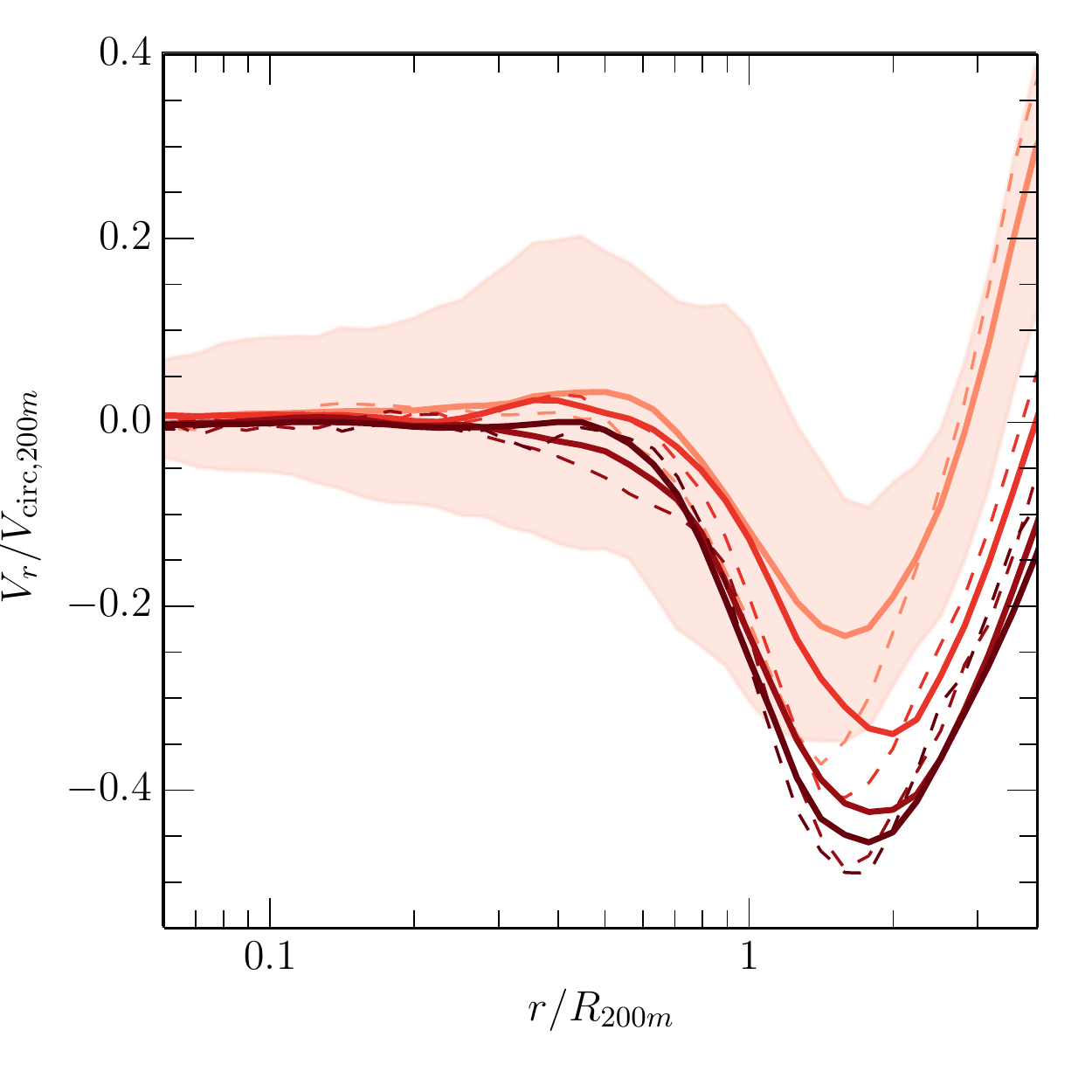}
\includegraphics[scale=0.65]{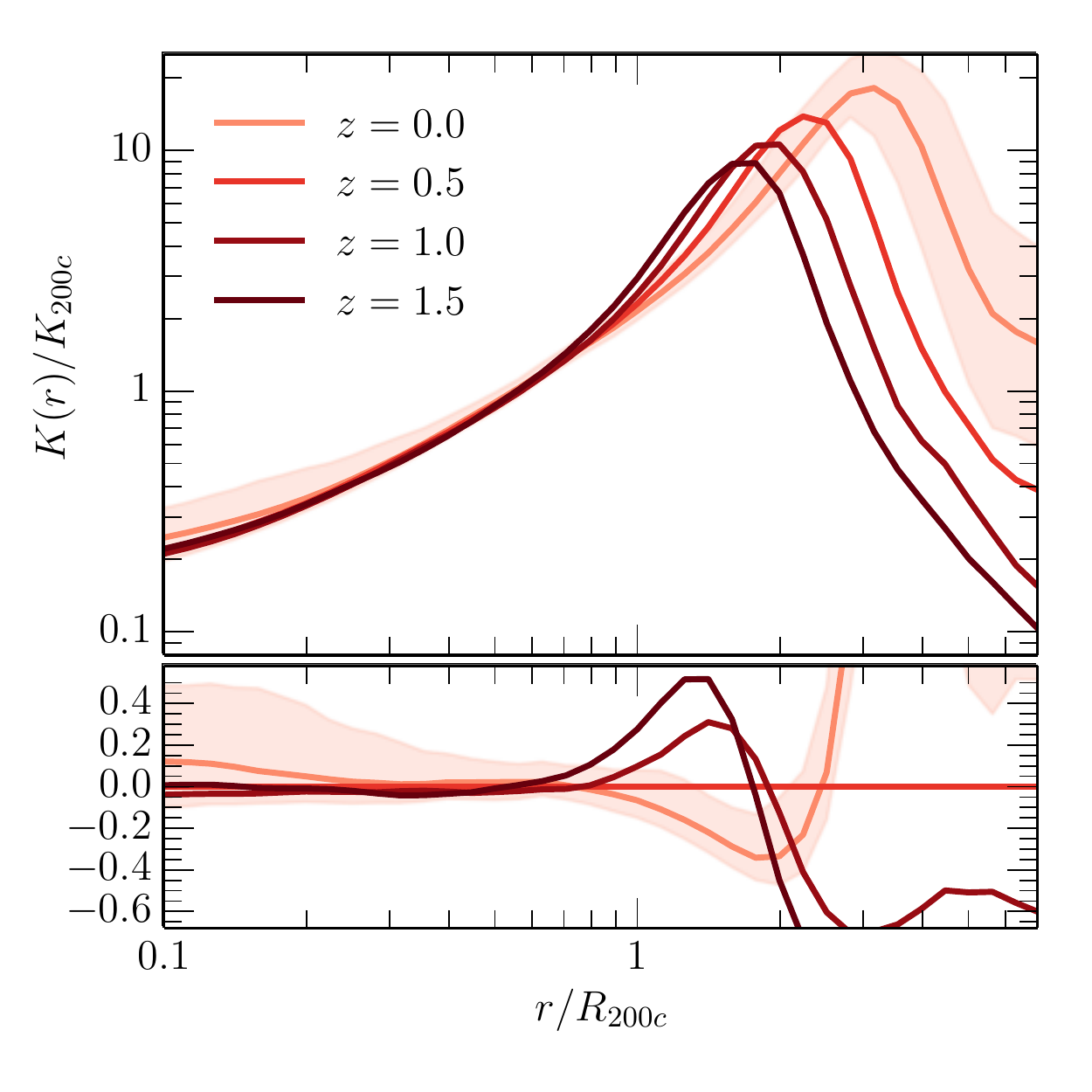}
\includegraphics[scale=0.65]{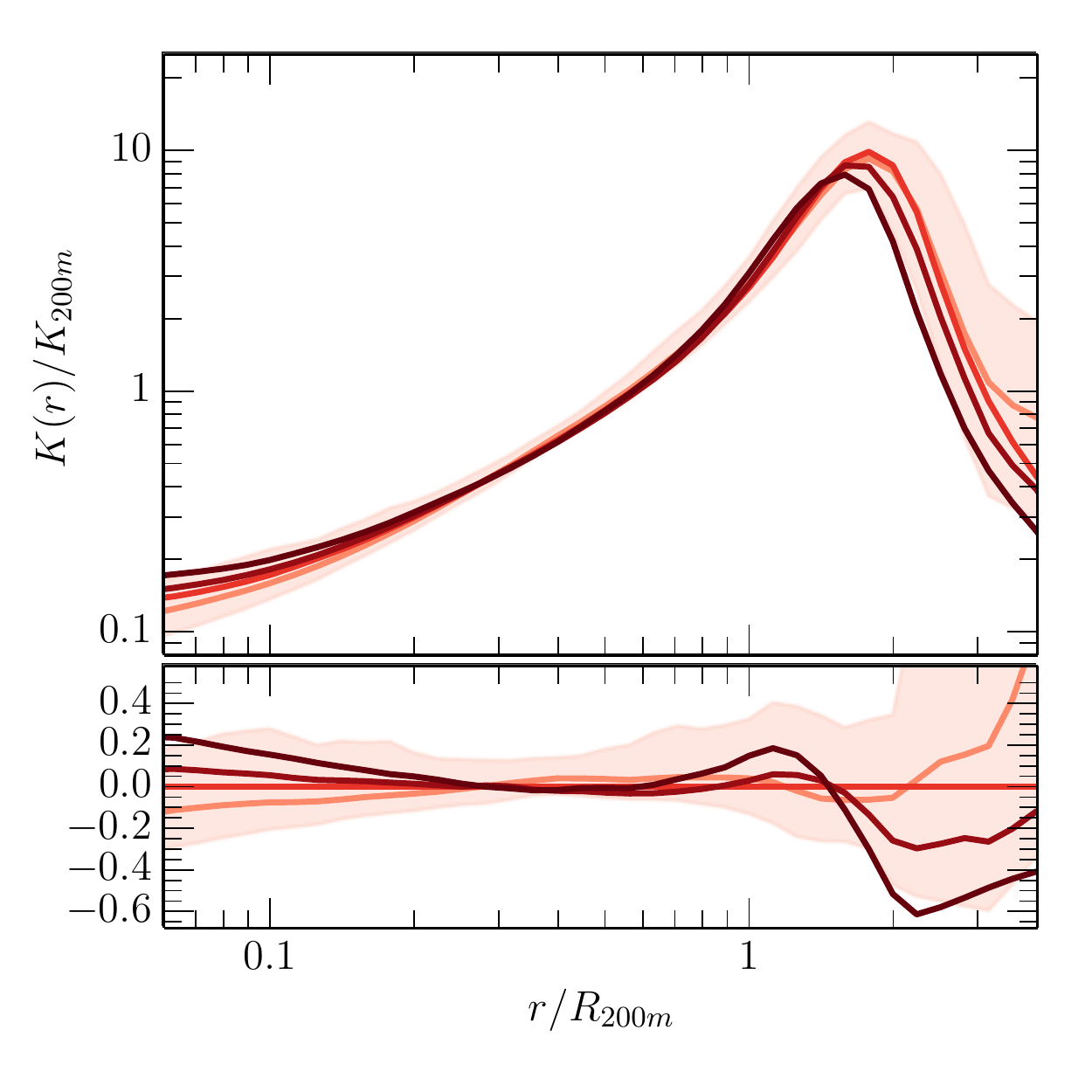}
\caption
{Profiles of average radial velocity ({\em top panels}) and  gas entropy ({\em bottom panels}) at $z=0, 0.5, 1.0,$ and $1.5$. The lower subplot in each panel shows the deviations of the profiles from different $z$ outputs relative to that of the $z=0.5$ clusters. The {\em left} panels show the profiles of cluster halos defined using the critical density, while the {\em right} panels show the profiles of the same cluster halos defined using the mean density. In the upper panel, we show the {\em mean} radial velocity profiles for gas and DM in {\em solid} and {\em dashed} lines respectively. The shaded regions indicate the $1\sigma$ scatter around the mean gas profiles for the $z=0$ clusters. }
\label{fig:ent_vr_r200c_r200m_z}
\end{center}
\end{figure*}

\subsection{Substructure Removal and Profile Making}
\label{sec:substructures}

In this work we are primarily interested in the thermodynamic properties of the diffuse ICM.
Gas in dense clumps and filaments is expected to have different thermodynamical properties from 
the diffuse ICM. In this work we minimize their effects by identifying and removing these gaseous 
substructures directly in simulations.  
Using the method proposed by \citet{zhuravleva_etal2013}, we remove gas clumps by 
excluding gas cells that have logarithmic density $3.5\sigma$ above the median for a given radial bin.  
Similarly we remove gaseous filamentary structures by excluding radially infalling gas cells with 
logarithmic density between $1\sigma$ and $3.5\sigma$ above the median for a given radial bin. 
As shown in Figure~\ref{fig:cluster_map}, this method can identify gas associated with 
clumps and filaments in the cluster outskirts quite well. 

After removing clumps and filaments, we compute the spherically averaged profiles 
by dividing the analysis region for each cluster into 99 spherical logarithmically 
spaced bins from $10\,h^{-1}{\rm kpc}$ to $10 \,h^{-1}{\rm Mpc}$ (comoving) 
in the radial direction from the cluster center, which is defined as the position with 
the maximum binding energy. Our results are insensitive to the exact choice of binning.  
We then compute volume-weighted mean profiles of density, pressure, and entropy, 
and mass-weighted mean profiles of gas temperature and gas velocities for each cluster halo, 
and present their mean profiles averaged the over the cluster sample. For the rest of the paper, 
all of the gas profiles are presented with substructures and filaments removed, unless noted otherwise. 

\section{Results}
\label{sec:results}

\subsection{Self-similarity of Gas Profiles}
\label{sec:redshift_dependence}

\begin{figure*}
\begin{center}
\includegraphics[scale=0.468]{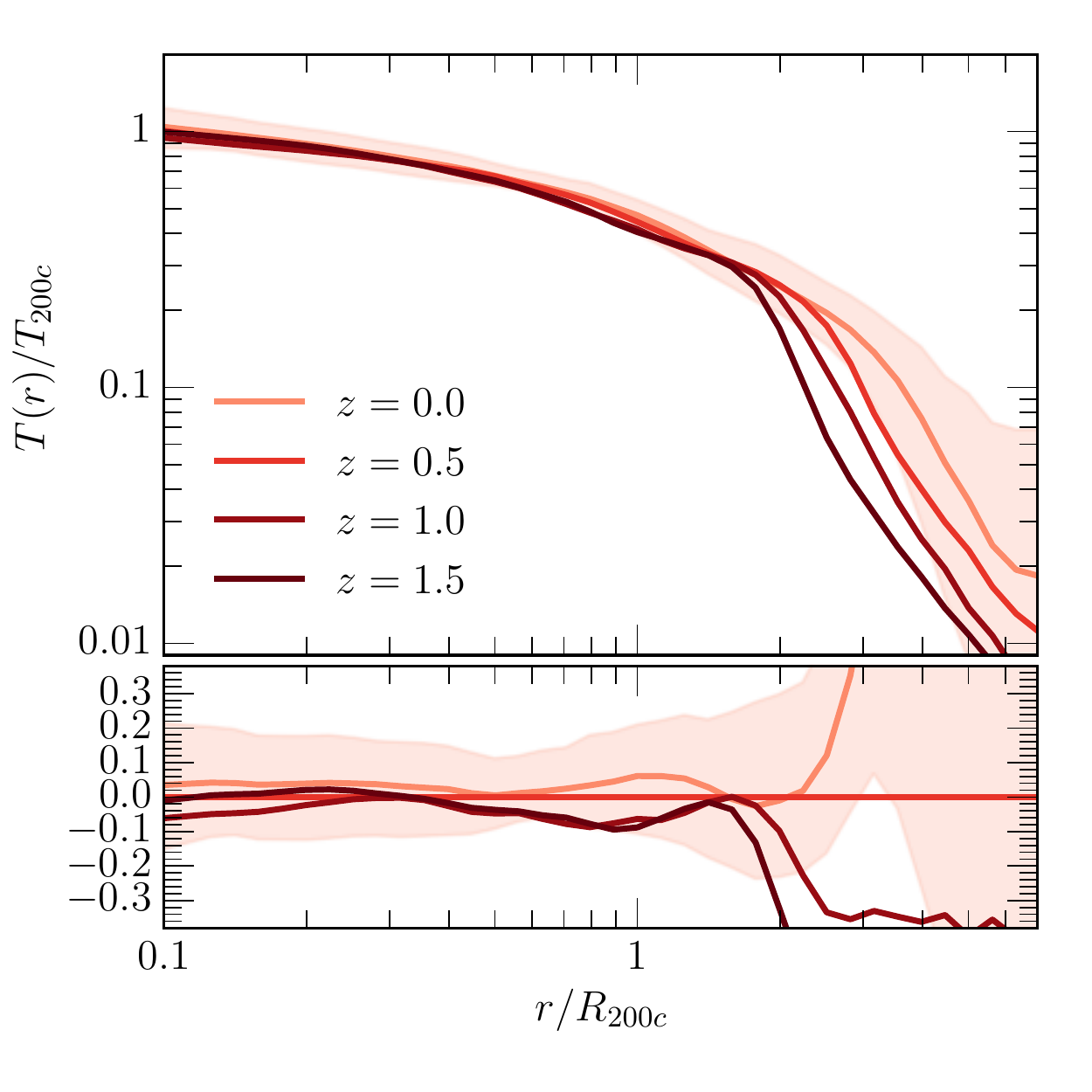}
\includegraphics[scale=0.468]{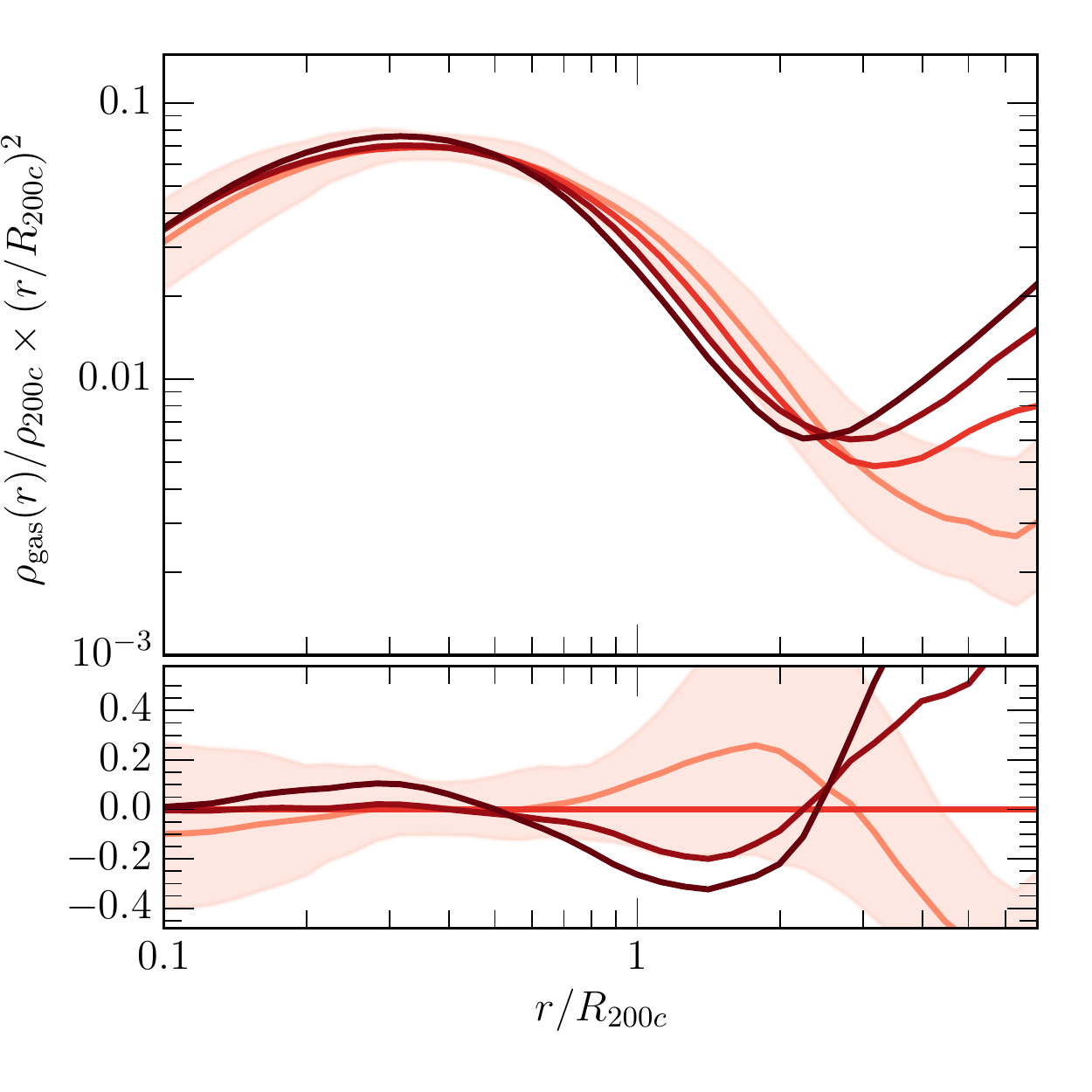}
\includegraphics[scale=0.468]{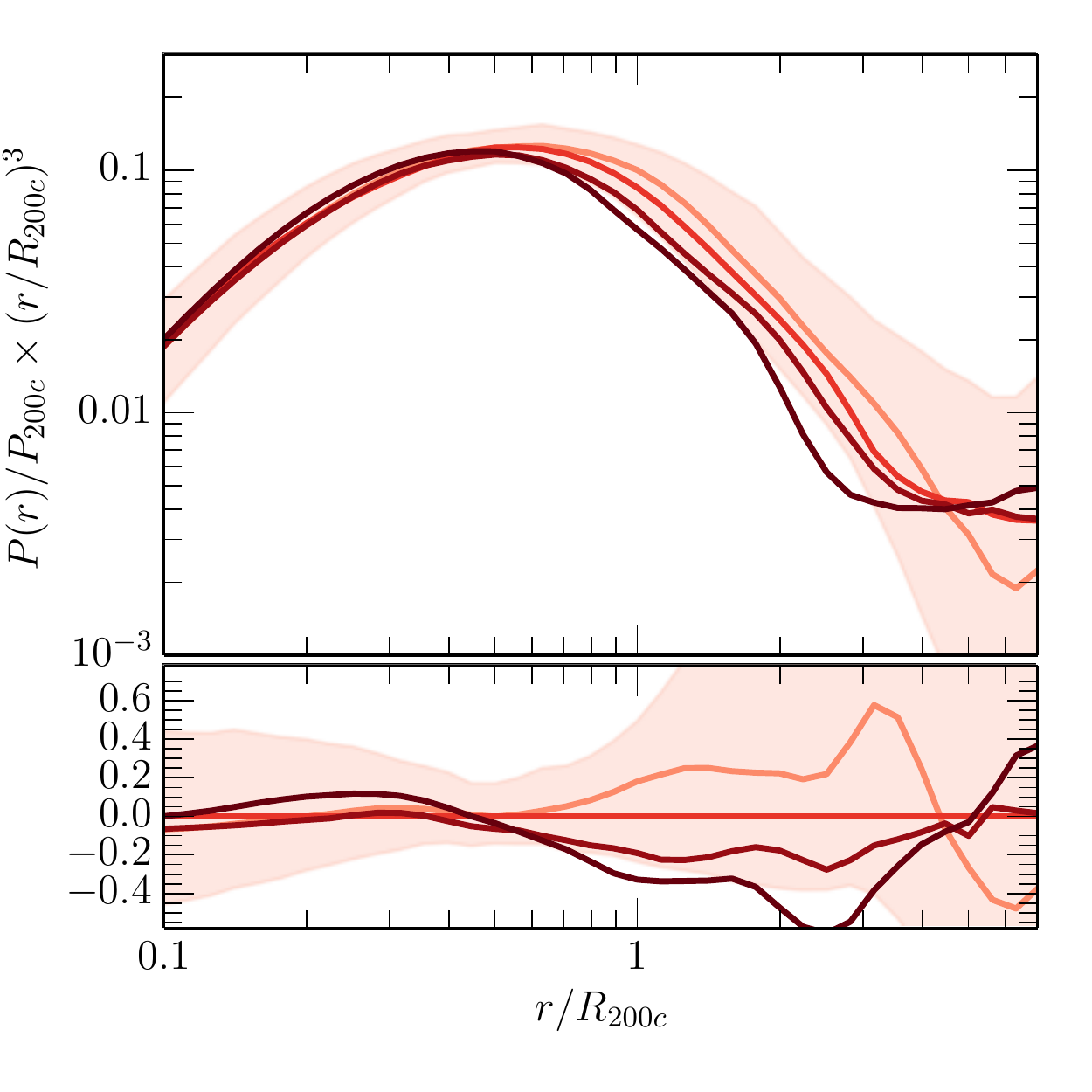}
\includegraphics[scale=0.468]{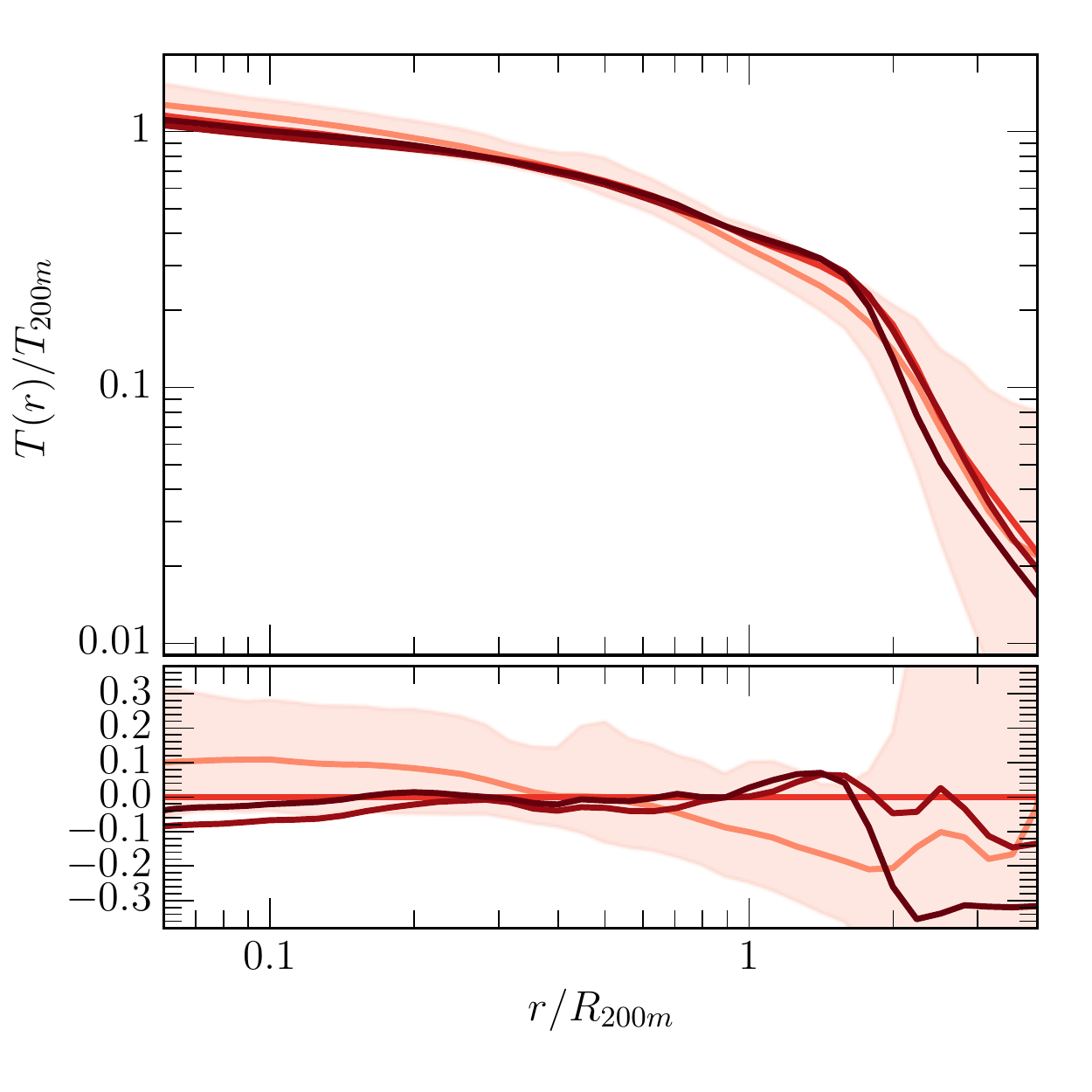}
\includegraphics[scale=0.468]{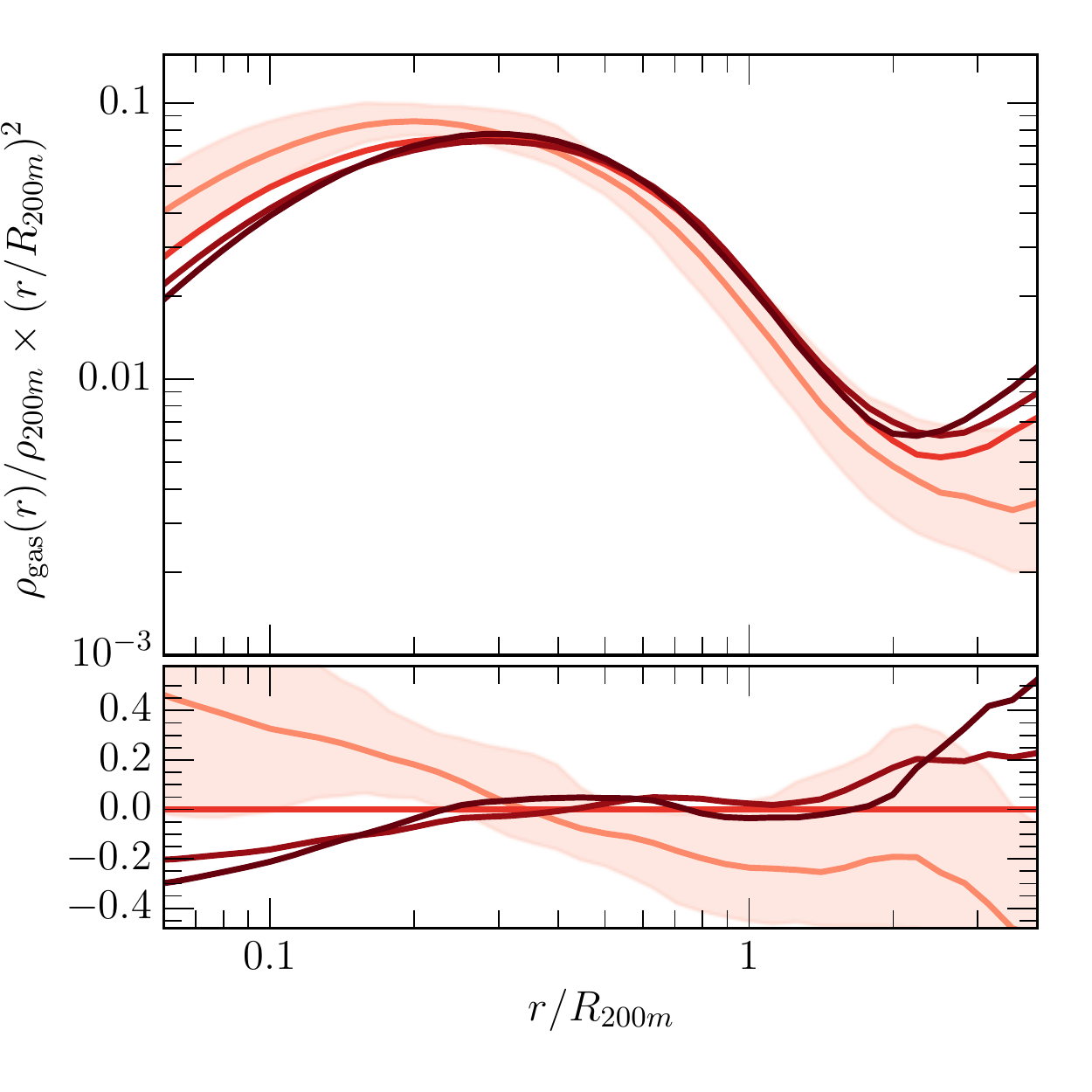}
\includegraphics[scale=0.468]{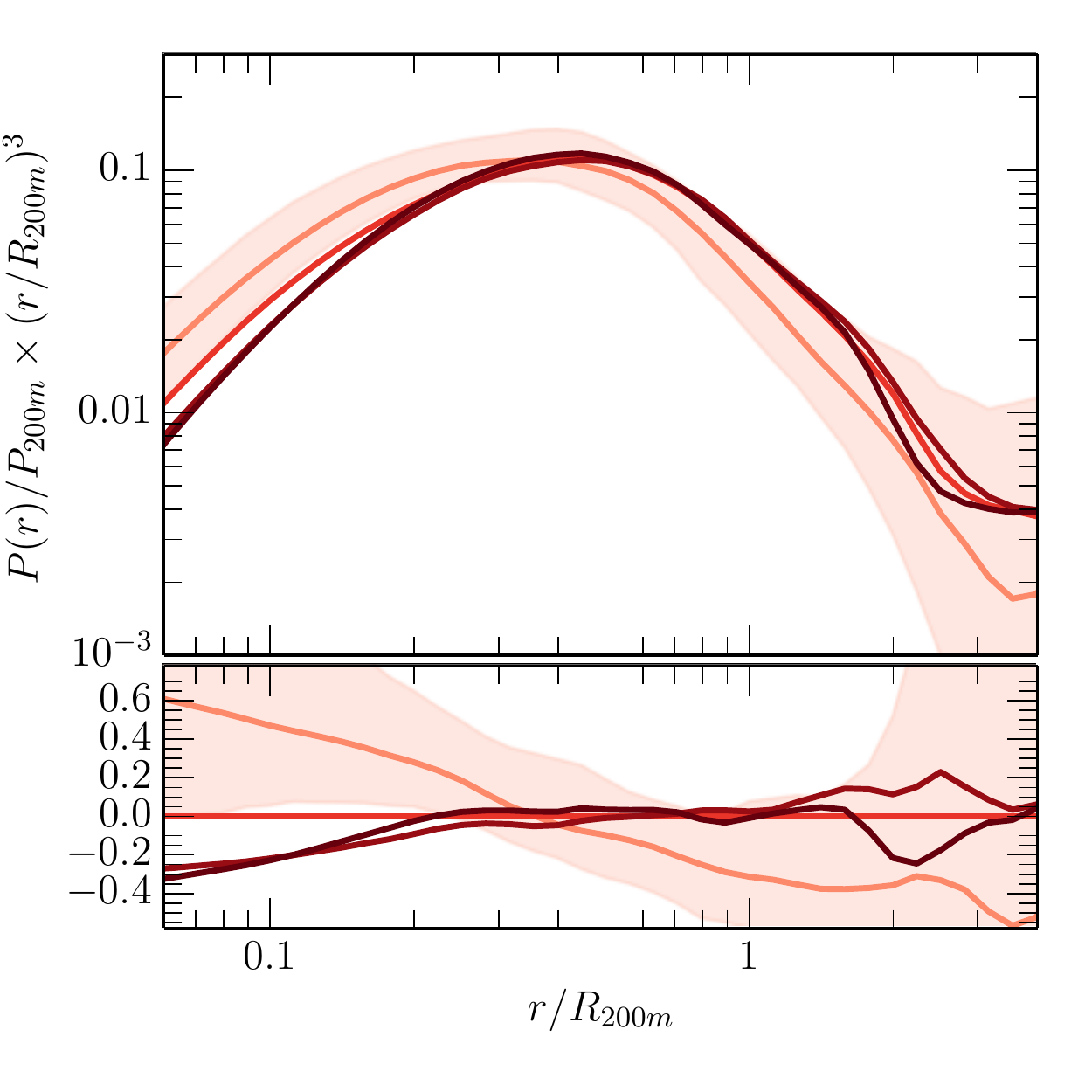}
 \caption
{From left to right, we show the profiles of gas temperature, density, and pressure for the cluster halos at $z=0.0, 0.5, 1.0,1.5$. The upper panels show the profiles normalized using the critical density $200\times \rho_c(z)$, while the lower panels show the same profiles normalized using the mean density $200\times \rho_m(z)$. In each figure, the bottom sub-panel shows the fractional deviations of the profiles with respect to the $z=0.5$ clusters. The shaded regions indicate the $1\sigma$ scatter around the mean for the $z=0$ clusters. }
 \label{fig:rho_temp_pres_r200c_r200m_z}
 \end{center}
 \end{figure*}

\begin{figure*}[t]
\begin{center}
\includegraphics[scale=0.65]{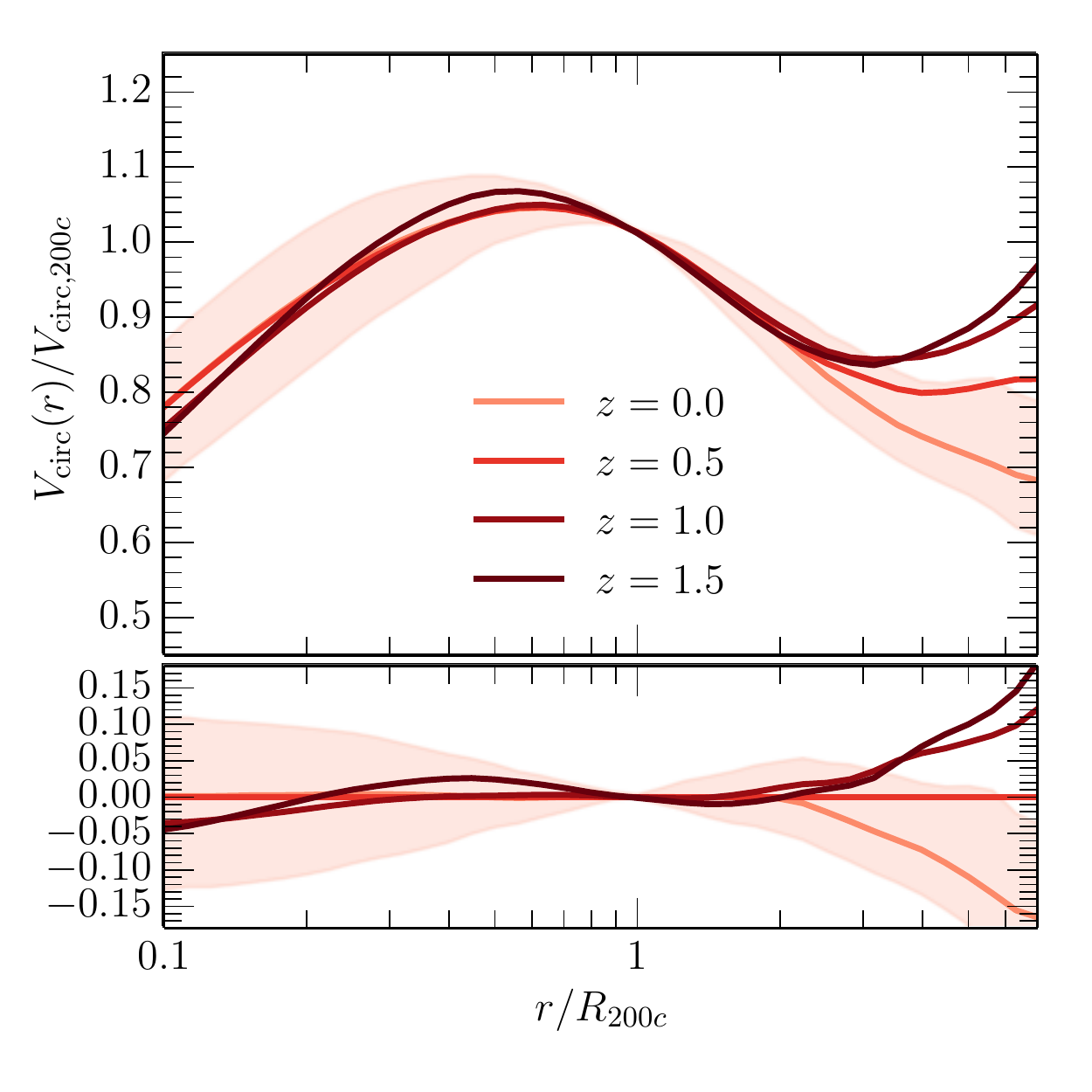}
\includegraphics[scale=0.65]{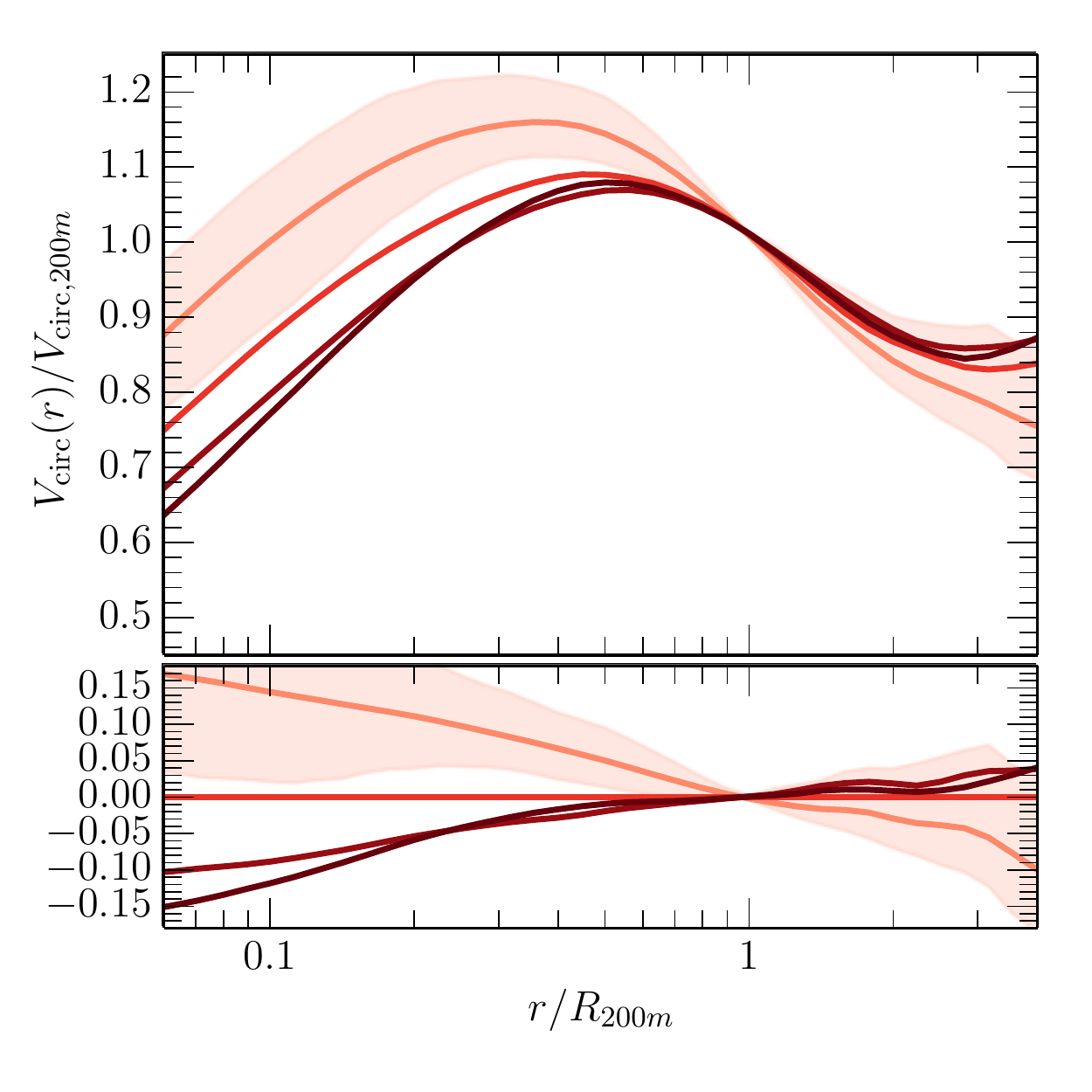}
\caption
{Profiles of circular velocity $V_{\rm circ} \equiv \sqrt{GM(<r)/r}$ at $z=0.0, 0.5, 1.0,1.5$. The {\em left} panel shows the profiles of the cluster halos defined using the critical density, while the {\em right} panel shows the same profiles with cluster halos defined using the mean density.   In each figure, the bottom sub-panel shows the fractional deviations of the profiles with respect to the profile of the $z=0.5$ clusters. The shaded regions indicate the $1\sigma$ scatter around the mean for the $z=0$ clusters. }
\label{fig:vcirc_r200c_r200m_z}
\end{center}
\end{figure*}

We begin by comparing radial velocity profiles of gas and DM in clusters defined 
with respect to the critical density and the mean density of the universe.  
The velocity profiles are normalized by the circular velocity 
$V_{{\rm circ}, \Delta} \equiv \sqrt{GM_{\Delta}/R_{\Delta}}$.  
In Figure~\ref{fig:ent_vr_r200c_r200m_z} we show the evolution of 
radial velocities for gas between $z=1.5$ and $z=0.0$.  
The top left panel shows the profiles for clusters normalized 
by using $\Delta_c=200$, and the top right panel shows the profiles
of the same clusters normalized by using $\Delta_m=200$. 
For ${\Delta_c}=200$, there is significant redshift evolution in the 
radial gas velocity profile, where $V_r (r= R_{200c})$ 
varies from $-0.3V_{200c}$ to $\approx 0$ from $z=1.5$ to $z=0$. 
Outside $R_{200c}$,  the radial velocities of DM and gas are mostly negative, 
indicating infall of DM and gas onto the cluster halos. 
The magnitude of the velocities in these infall regions evolves with redshift: 
higher-$z$ clusters show more negative radial velocities, 
indicating their higher MAR. The location of the velocity minimum,
where the gas is infalling most strongly, also evolves with $z$, 
and it is located at larger fraction of $R_{200c}$ at lower $z$. 
In the same figures, we also show the average radial velocities of DM 
(indicated by the dashed lines).  
Although the radial velocity profiles of both gas and DM show 
similar qualitative trend with $z$, gas is generally accreting at a slower rate than DM; 
the collisional gas\footnote{Note that the mean free path of electrons in the ICM plasma 
could be large ($\gtrsim 100 \kpc$), which may break down 
the hydrodynamic approximation in cluster outskirts. 
However, the presence of any magnetic field may reduce the effective mean free path 
below the numerical resolution such that the ICM can be treated effectively 
as collisional. This uncertainty must be kept in mind when interpreting our results. } 
experiences ram pressure from the surrounding ICM, while the collisionless DM does not. 

At large radii ($r \gtrsim 3R_{200c}$), gas traces DM, where the velocities of both components 
follow the Hubble flow in the expanding universe.  Their normalized radial velocity profiles are 
independent of redshift, because the normalization with respect to $R_{200c}$ and $V_{\rm circ,200c}$ 
naturally accounts for the Hubble flow: $R_{200c} \propto \rho_c^{-1/3} \propto H(z)^{-2/3}$, and 
$V_{\rm circ,200c} \propto H(z)^{1/3}$, so $H(z)=V_r / r \propto  V_{\rm circ,200c}/R_{200c}$,
leaving the normalized radially velocity profile $V_r(r/R_{200c})/V_{\rm circ,200c}$ redshift independent. 

An interesting radius is the turnaround radius $R_\ta$, where the radial velocity becomes 
zero as the accreting mass detaches from the Hubble flow. The top left panel of
Figure~\ref{fig:ent_vr_r200c_r200m_z} shows that $R_\ta/R_{200c}$ 
(the outermost radius where the radial velocity is zero) is independent 
of redshift, and it is located at $R_\ta/ R_{200c} \approx 5$. We expect $R_\ta$ to
follow the evolution of the Hubble parameter, as it is the radius where the dynamics of both
dark matter and gas detach from the Hubble flow. Therefore $R_\ta$ is well-traced by 
$R_{200c}$ which is defined in terms of the Hubble parameter.

On the other hand, $R_\ta$ does not scale well with $R_{200m}$.
The radial velocity profiles $V_r/V_{\rm circ,200m}$ outside the infall region
show strong evolution with $z$, especially at late times ($z\lesssim 1.0$). 
This is because $R_{200m} \propto \rho_m(z)^{-1/3}$ only accounts for the 
evolution of the matter component, but does not account for the effects of dark energy 
which drives the accelerated expansion of the universe at low $z$, leading to strong
evolution trend in $R_\ta/R_{200m}$ and the radial velocity profile outside the infall region. 

However, at smaller radii, we find that choosing ${\Delta_m}=200$ 
makes the radial velocity profile more universal with $z$. The locations of radial velocity 
minima in both DM and gas for ${\Delta_m}=200$
show little redshift evolution compared to the case of ${\Delta_c}=200$. 
This behavior is expected, as the accreting matter undergoes free-fall 
once decoupled from the Hubble flow, where its evolution is governed primarily 
by the gravity inside the virialized region, whose mass density is 
characterized by the cosmic mean mass density independent of the Hubble parameter. 

Similar trends are observed in the entropy profiles of gas. 
The profiles are normalized by the self-similar values described in the
Section~\ref{sec:self-similar-collapse}.  
The bottom panels of Figure~\ref{fig:ent_vr_r200c_r200m_z} 
show the redshift evolution of the entropy profiles. 
Each line represents the entropy profile averaged over the cluster sample 
at $z=0.0, 0.5, 1.0,1.5$. Each profile shows a well-defined entropy peak, 
which corresponds to the location of the accretion shock\footnote{We define accretion shocks as 
regions where pristine gas from voids falls into the cluster and gets shock-heated for the first time, 
which are commonly referred to as ``external'' shocks, in contrast to ``internal'' shocks driven 
by mergers or accretions through filaments \citep[e.g.,][]{ryu_etal2003}.}, 
$R_{\rm sh}$, where the gas radial velocity is minimum. 

Some care is needed when interpreting the accretion shock radius $R_{\rm sh}$ defined using 
the entropy peak or the minimum of the radial infall velocity of gas. 
When the accreting collisional gas is shocked, it stops infalling and its radial velocity should 
jumps abruptly to zero. Likewise, the gas entropy profile should jump sharply behind the accretion shock when 
the gas is heated through the shock.  However, these sharp jumps are not seen in our radial 
velocity nor entropy profiles, which increase more smoothly in the post-shock region.
This is partly because gas accretion is aspherical, and the actual topology of the accretion 
shocks is rather complex 
\citep[e.g.,][]{ryu_etal2003, pfrommer_etal2006, skillman_etal2008,vazza_etal2009b,planelles_quilis2013,schaal_springel2014}.  
Spherically averaging the velocity and entropy profiles will smooth out these jumps. 
Moreover, cold gas accreting along filaments can penetrate deeper into the cluster 
and are shocked at smaller cluster-centric radii \citep[e.g.,][]{molnar_etal2009}. 
This creates a series of shocks with varying velocity and entropy jumps,
smoothing the velocity and entropy profiles in the infall region. 

For $\Delta_c=200$, the accretion shock systematically increases toward larger 
cluster-centric radii at lower $z$.  The entropy measured at $r=R_{200c}$ 
decreases from $3\times K_{200c}$ at $z=1.5$ to $1.5\times K_{200c}$ at $z=0$ 
due to the combination of the shift in the location of the entropy peak and 
the evolution in the entropy normalization $K_{200c}$.  On the other hand, 
for $\Delta_m=200$, the entropy profiles in the radial range of 
$0.3 \leq r/R_{200m} \leq 1.0$ remain roughly constant with $z$. 
The accretion shock radius $R_{\rm sh}/R_{200m}$ does not evolve much, 
and it is located at $R_{\rm sh}\approx 1.6 R_{200m}$ at all $z$. 

In Figure~\ref{fig:rho_temp_pres_r200c_r200m_z}, 
we show profiles of other thermodynamic quantities: 
gas density, temperature and thermal pressure profiles normalized 
using $\Delta_c=200$ and $\Delta_m=200$ at different $z$. 
We find that these ICM profiles exhibit similar $z$-dependence 
to that of the entropy profile. For $\Delta_c=200$, there are systematic 
deviations of the profiles with $z$ outside the cluster core ($r \ge 0.2R_{200c}$).  
For example at $r=R_{200c}$, high-$z$ clusters have lower normalized 
temperature because they have higher physical mass accretion rate, 
so that their accretion shocks can penetrate deeper by pushing 
the low-temperature pre-shock regions toward the inner regions of clusters.
Similarly, the evolution in the normalized density and pressure profiles 
is also due to the evolution in $R_{\rm sh}/R_{200c}$. 
On the other hand, switching the halo definition to $\Delta_m=200$ 
captures the redshift evolution of the $R_{\rm sh}$ much better and 
results in thermodynamic profiles that are more universal 
with $z$ in the radial range of $0.3 \le r/R_{200m} \le 1.0$. 

The self-similar secondary infall model predicts that the location of shock
is located at a fixed fraction of the current turnaround radius, with $R_{\rm sh} =
0.347 R_\ta$, in the flat matter-dominated universe \citep{bertschinger1985}. 
However, our simulation shows that $R_{\rm sh}$ is {\em not} proportional to $R_\ta$. 
$R_{\rm sh}$ evolves as $R_{200m}$, whereas $R_\ta$ evolves as $R_{200c}$. 
This deviation from the self-similar secondary infall model is likely due to the increasing
effects of dark energy at low $z$, which breaks the self-similar evolution of infalling matter
that is currently turning around. We will investigate the origin of this deviation in a future paper. 

We find little dependence of the gas profiles on cluster mass or density peak height 
$\nu \equiv \delta_c/\sigma(M,z)$ defined in Section~\ref{sec:self-similar-collapse}, because 
our sample contains only massive clusters. On average, halos with higher mass or peak height 
are on average accreting more rapidly compared to those with low mass or peak height 
\citep{cuesta_etal2008, mcbride_etal2009, diemer_kravtsov2014}, 
which can introduce mass or peak height dependence in the outskirt gas profiles. 

Baryonic processes such as radiative cooling, star formation, and
energy feedback from supernovae or active galactic nuclei can
significantly influence the gas profiles in the inner regions and
break self-similar behavior.  These physical processes can also change
the thermodynamical structure at larger radii and influence gas
dynamics in low mass halos, such as galaxies and galaxy groups
\citep[e.g.,][]{faucher-giguere_etal2011,vandevoort_etal2011}.  We
expect the effects of baryonic physics to be considerably smaller in
the outskirts of massive halos, where gravitational physics dominates
the gas dynamics.  However, further work is necessary to fully
quantify how baryons affect the MAR and the self-similarity of gas
profiles in all halos.  A study of gas flows in group and galaxy-size
halos help address these issues \citep[e.g.,][]{wetzel_nagai2014}.

\subsection{Evolution of the Circular Velocity Profile}
\label{sec:vcirc}

The differences in the self-similarity between the inner and outer regions can be understood 
in terms of the evolution of the gravitational potential well,  
which determines the thermodynamical properties of the cluster gas. 
In Figure~\ref{fig:vcirc_r200c_r200m_z}, we plot the circular velocity profile $V_{\rm circ} (r)$
 which we use as a proxy for the gravitational potential, normalized using $\Delta_c = 200$ 
 and $\Delta_m = 200$ at $z=0.0, 0.5, 1.0,1.5$. For $\Delta_c = 200$, 
 the circular velocity profile is more universal with $z$ at  $r \le R_{200c}$, 
 indicating that the evolution of the cluster potential in the inner region is well captured 
 by $\rho_c(z)$. On the other hand for $\Delta_m = 200$, $V_{\rm circ}$ evolves 
 significantly at $r\le R_{200m}$, but exhibits an enhanced level of self-similarity at $r\ge R_{200m}$.

The reason behind the dependence is that the gravitational potential well of the cluster halo
is already set during the early stage of formation of the halo \citep{li_etal2007,vandenbosch_etal2014}, 
while the outer region is more sensitive to the recent mass 
growth of the halo. $N$-body simulations have shown that halo grows in two phases: 
an early fast growth phase when the halo is formed via violent relaxation and phase mixing, 
followed by slow growth phase via smooth accretion \citep{wechsler_etal2002, zhao_etal2003}. 
The initial fast growth phase determines the inner density and hence the potential well of the halo. 
For a massive cluster-size halo at $z=0$, its fast growth phase occurs at high redshift ($z\gtrsim 1$) 
when the universe is still flat and matter-dominated, 
with its gravitational potential forming and scaling with the background density 
where $\rho_m(z)=\rho_c(z)$. 
During the subsequent slow growth phase in the 
epoch of dark energy domination ($\rho_m(z) < \rho_c(z)$), 
accretion only adds mass in the halo outskirts, leaving the inner mass distribution unchanged.
For constant mass, the radius defined with respect to the critical density scales as
$R_{\Delta_c} \propto \rho_c(z)^{-1/3} \propto E(z)^{-2/3}$, which evolves much slower than
$R_{\Delta_m} \propto (1+z)^{-1}$ at late times when the universe is no longer matter-dominated.  
Therefore, $R_{\Delta_c}$ tracks the slowly changing interior better, 
while $R_{\Delta_m}$ tracks the outer gas profiles determined by the mass accretion at late times. 
Further works are needed to understand the tight self-similar scaling of the inner profiles with $\rho_c(z)$. 
However, we note that in reality, the inner profiles are modified by baryonic physics 
which breaks the self-similarity \citep[e.g.,][]{mcdonald_etal2014}.

\begin{figure}
\begin{center}
\includegraphics[scale=0.65]{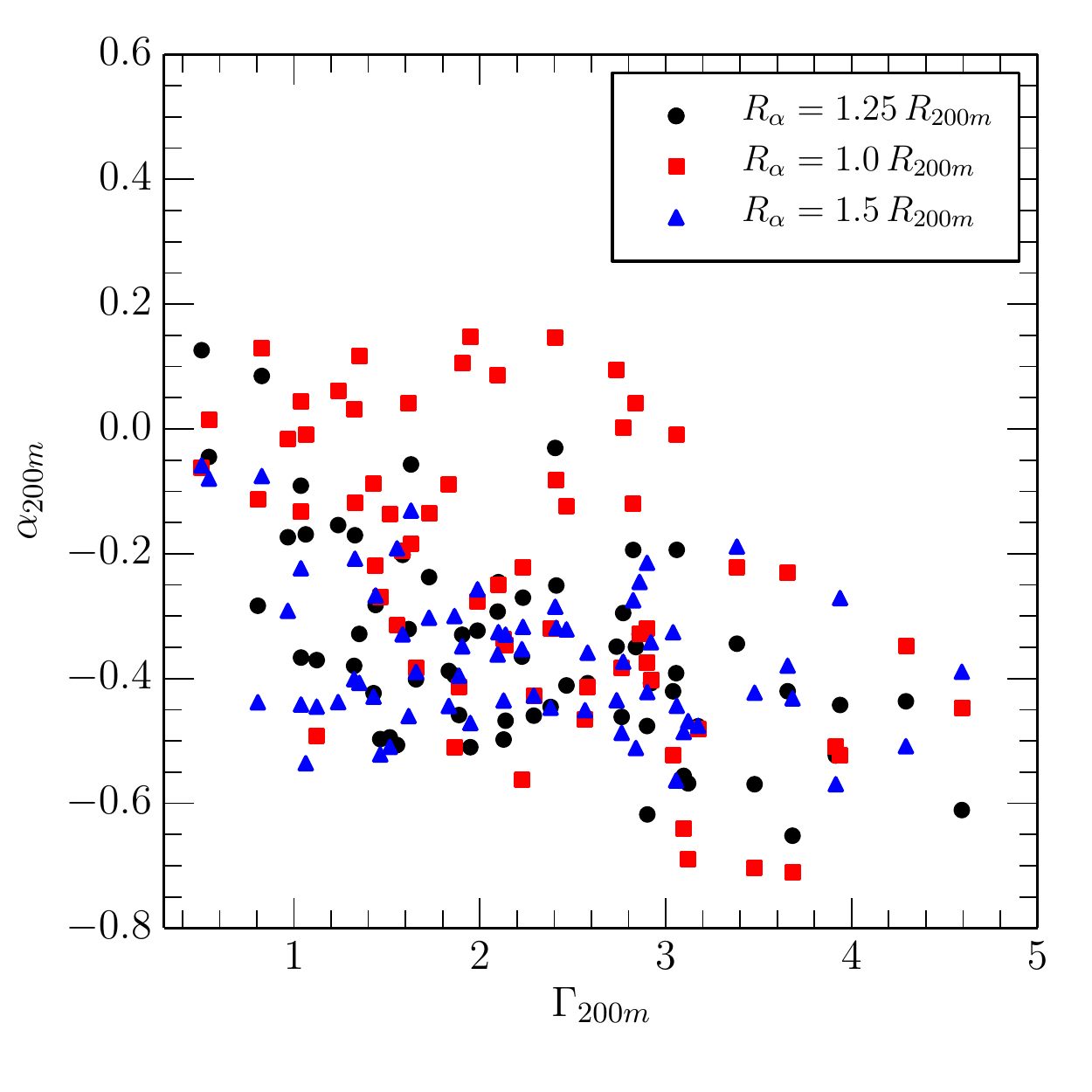}
\caption
{Comparison between two MAR proxies $\alpha_{200m}$ versus $\Gamma_{200m}$ for the $z=0$ clusters. Here, we consider three values of $\alpha_{200m}$ based on the DM radial velocity at $r=1.25 R_{200m}$ (black circles), $1.0 R_{200m}$ (red squares), and  $1.5 R_{200m}$ (blue triangles), normalized by the circular velocity at $r=R_{200m}$. }
\label{fig:gamma_alpha}
\end{center}
\end{figure}

\begin{figure}
\begin{center}
\includegraphics[scale=0.65]{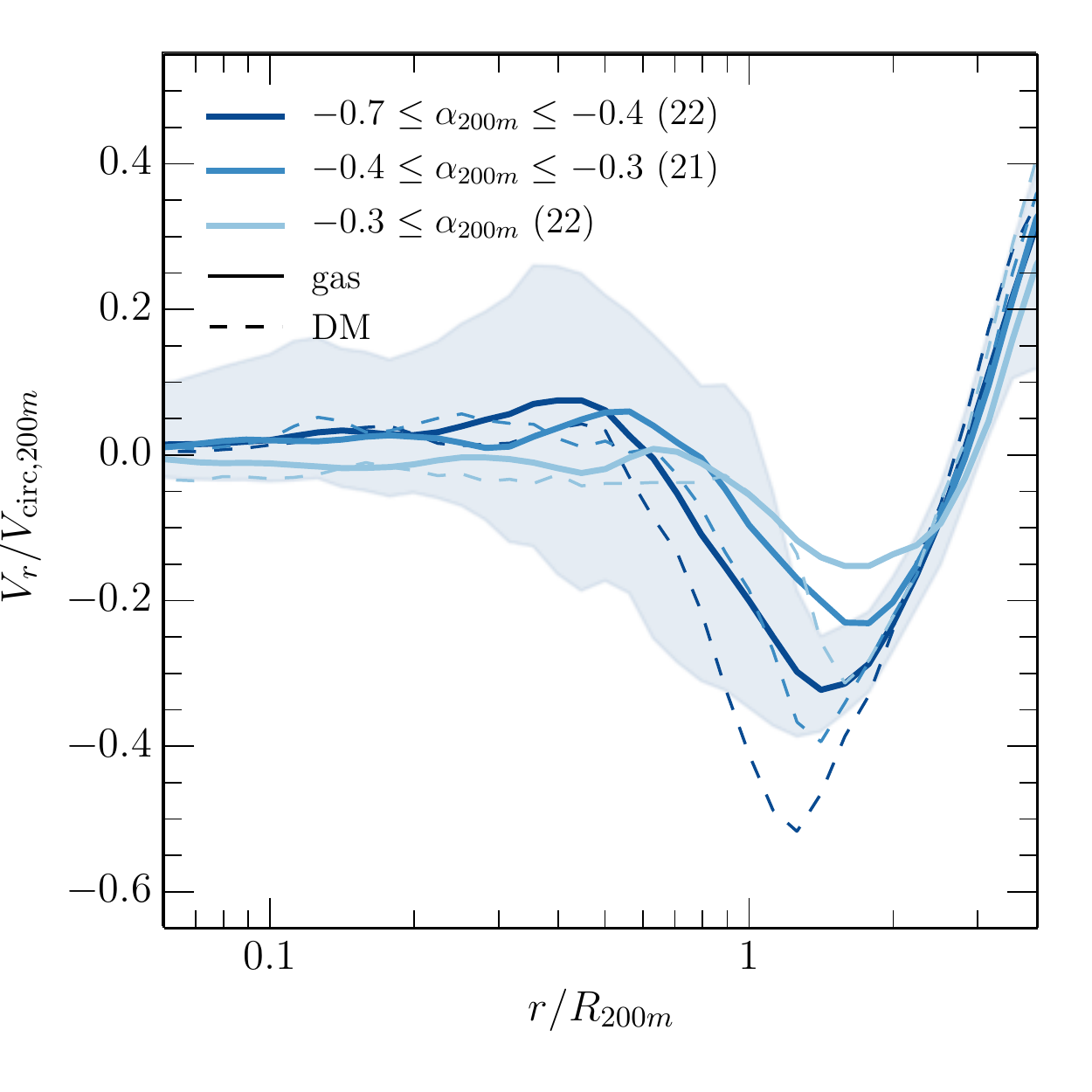}
\caption
{Profiles of radial velocity of DM ({\em dashed} lines)  and gas ({\em solid} lines) in different bins of MAR $\alpha_{200m}$ for the $z=0$ clusters.  The shaded regions represent 1$\sigma$ scatter in the profiles with the most negative $\alpha_{200m}$. The scatter for the other $\alpha_{200m}$ bins are similar in size.  The numbers in the parentheses indicate the number of clusters in each $\alpha_{200m}$ bin. }
\label{fig:vrad_alpha}
\end{center}
\end{figure}

\subsection{Effects of MAR on Gas Profiles}
\label{sec:mar_dependence}

\begin{figure*}
\begin{center}
\includegraphics[scale=0.468]{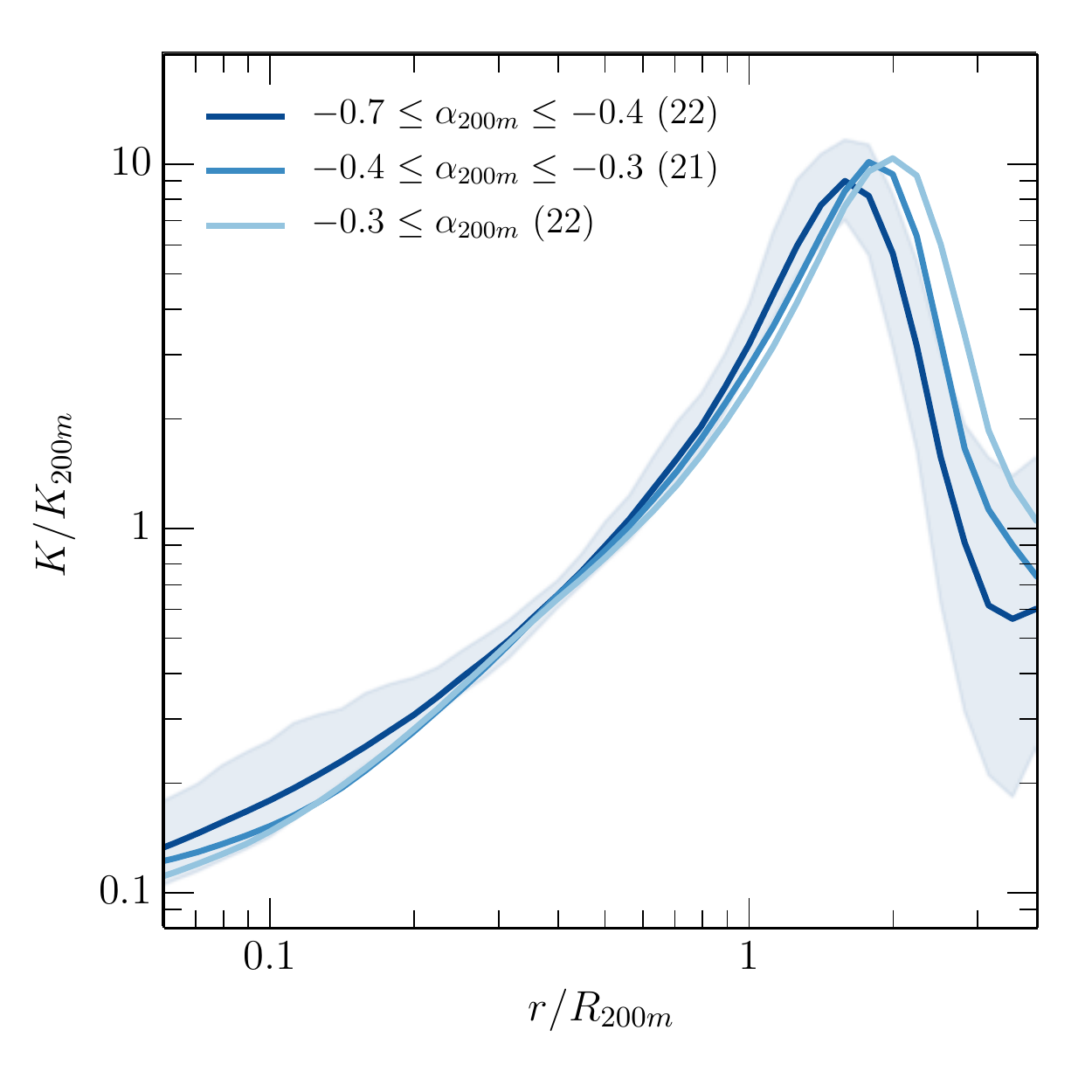}
\includegraphics[scale=0.468]{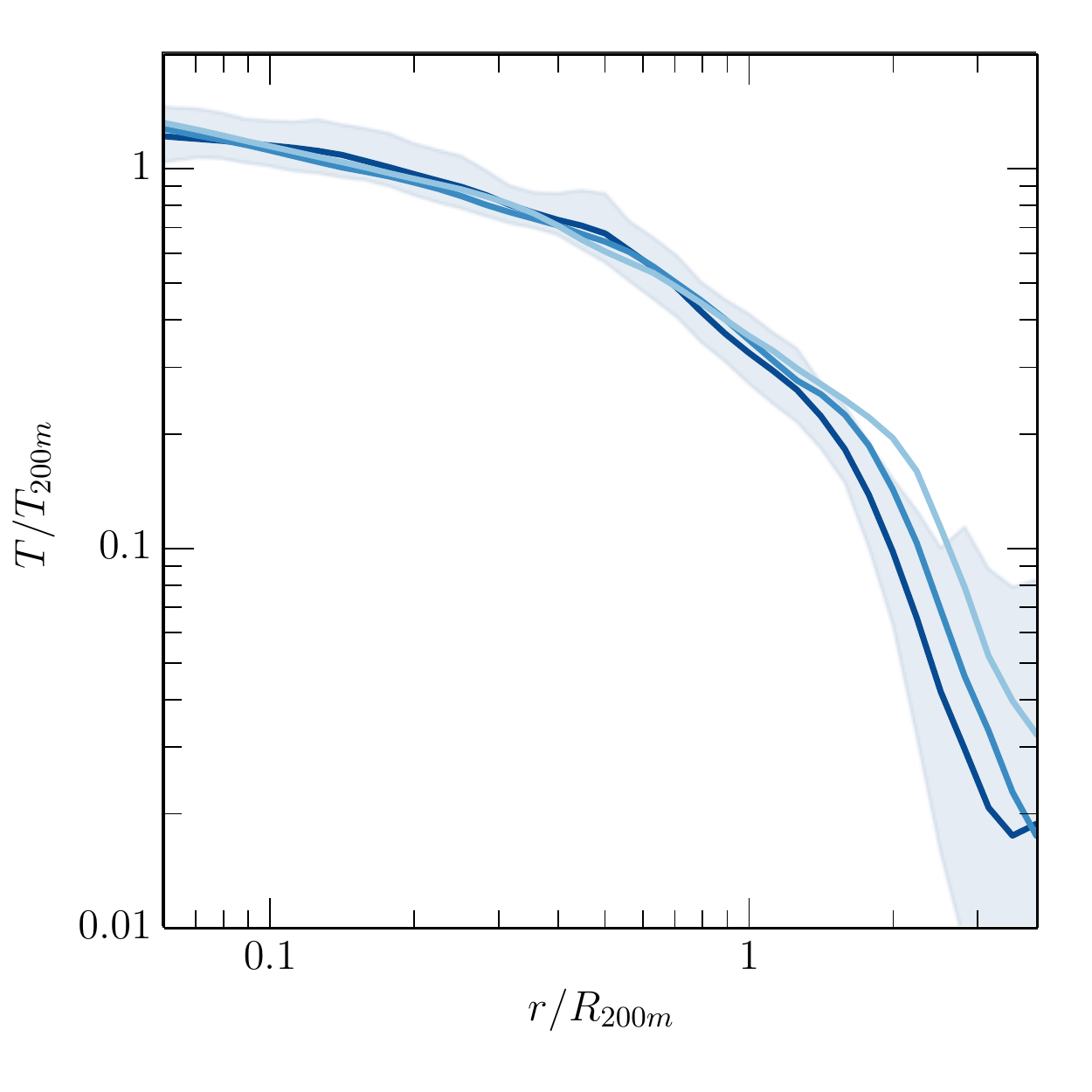}
\includegraphics[scale=0.468]{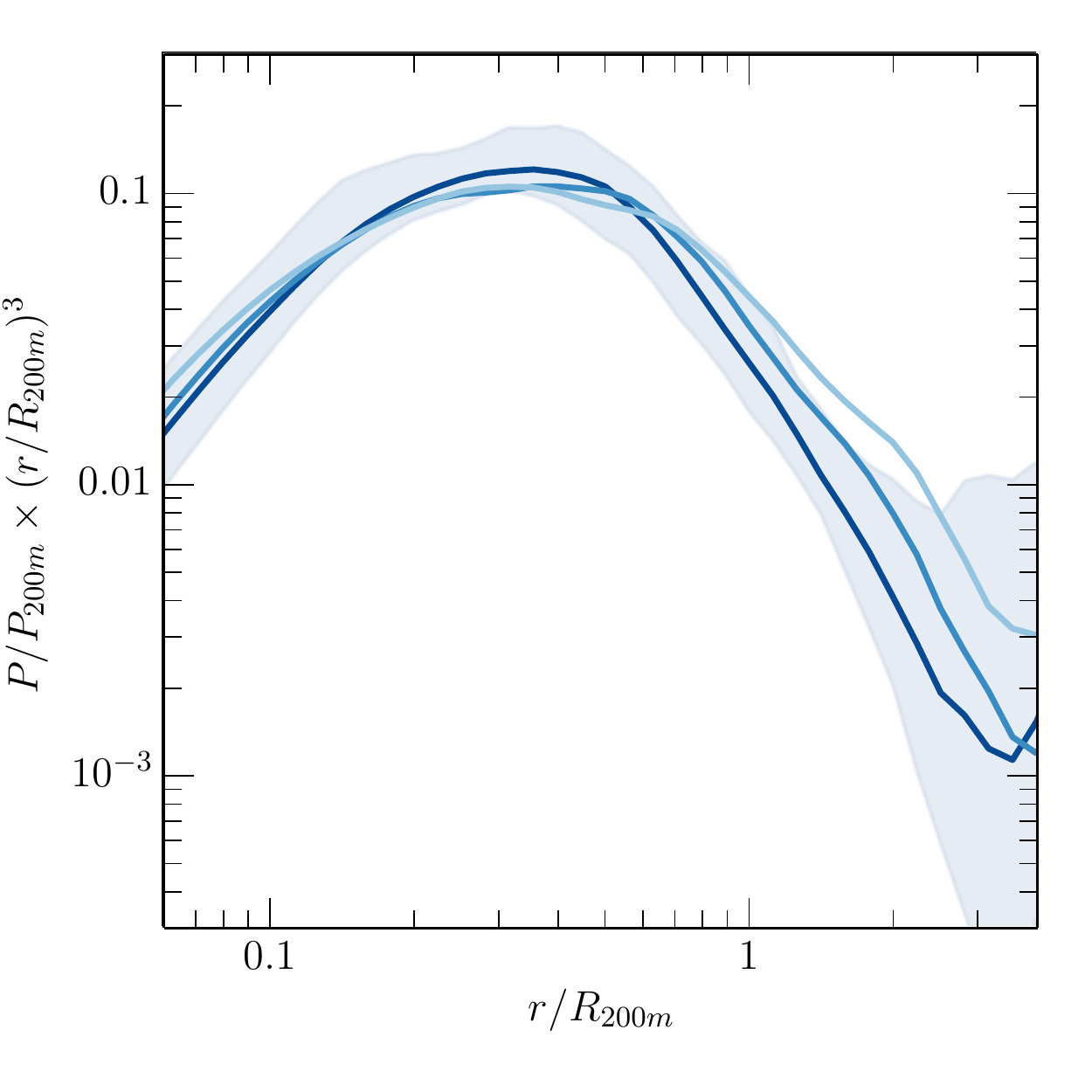}
\caption
{Profiles of gas entropy ({\em left}), temperature ({\em middle}),  and pressure ({\em right}) scaled with $R_{200m}$ as a function of MAR $\alpha_{200m}$ for the $z=0$ clusters.   We have multiplied the pressure profile by the radius cubed to show its dependence on $\alpha_{200m}$ more clearly. The shaded regions represent 1$\sigma$ scatter in the profiles with the most negative $\alpha_{200m}$. The scatter for the other $\alpha_{200m}$ bins are similar in size.  
The numbers in the parentheses indicate the number of clusters in each $\alpha_{200m}$ bin. 
}
\label{fig:ent_temp_pres_r200m_z0_vrad}
\end{center}
\end{figure*}

\begin{figure*}
\begin{center}
\includegraphics[scale=0.65]{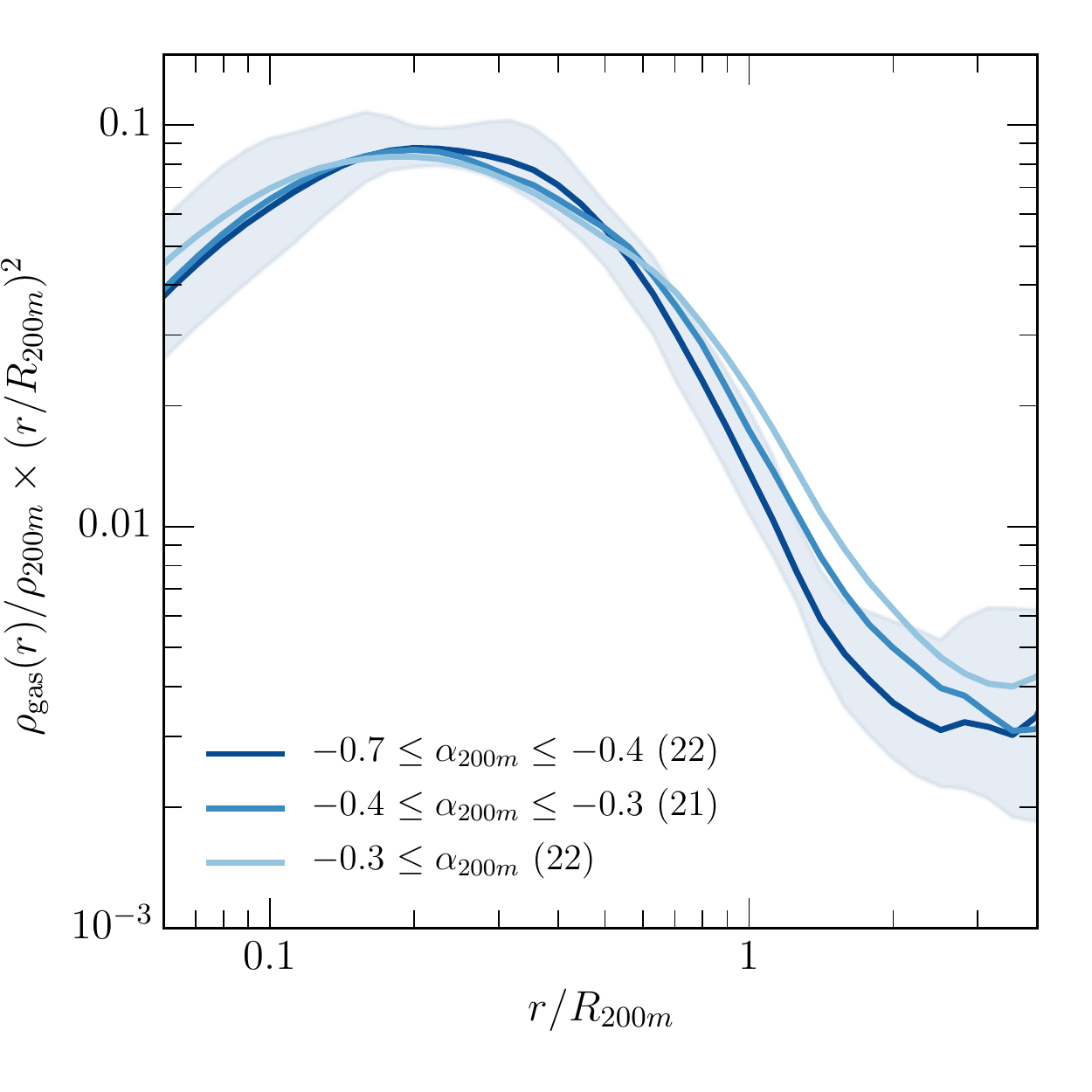}
\includegraphics[scale=0.65]{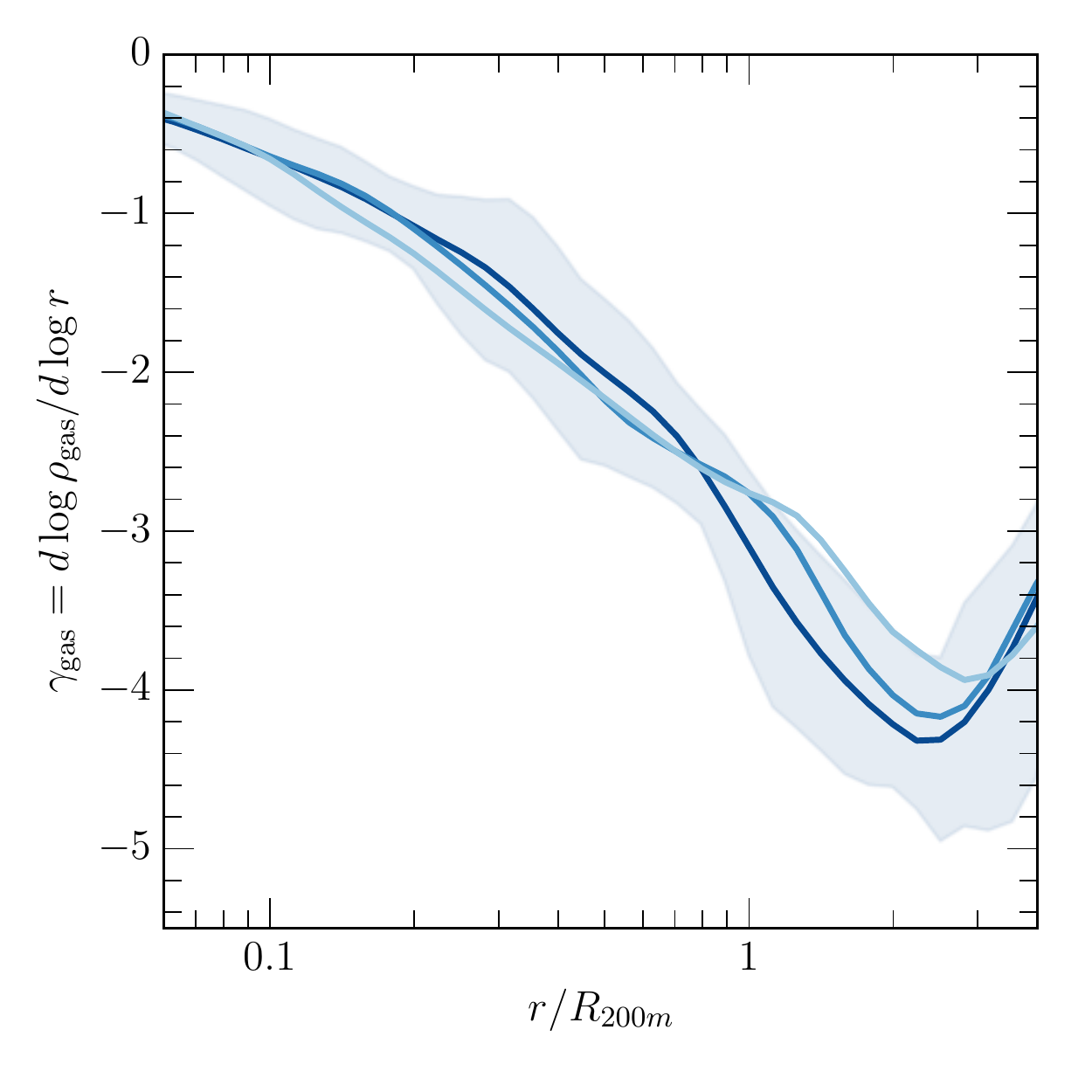}
\caption
{Profiles of gas density ({\em left}) and its logarithmic slope ({\em right}) as a function of the cluster-centric radius $r/R_{200m}$ for different MAR $\alpha_{200m}$ at $z=0$. The pressure profiles are multiplied by the square of the normalized radius to demonstrate its dependence on $\alpha_{200m}$ more clearly. The shaded regions represent 1$\sigma$ scatter in the profile with the most negative $\alpha_{200m}$. The scatter for the other $\alpha_{200m}$ bins are similar in size.  The numbers in the parentheses indicate the number of clusters in each $\alpha_{200m}$ bin. 
}
\label{fig:rho_r200m_z0_vrad}
\end{center}
\end{figure*}

\begin{figure*}
\begin{center}
\includegraphics[scale=0.65]{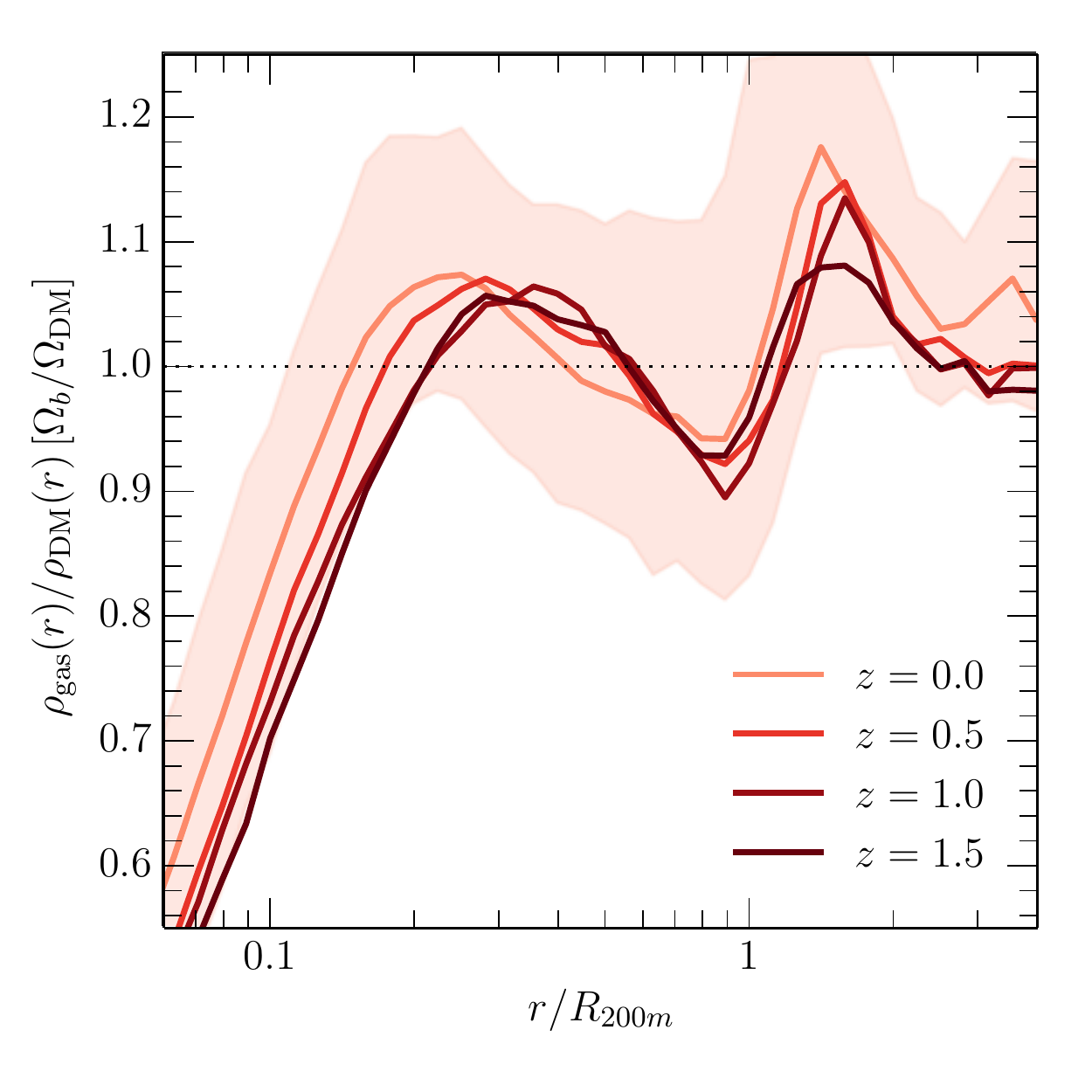}
\includegraphics[scale=0.65]{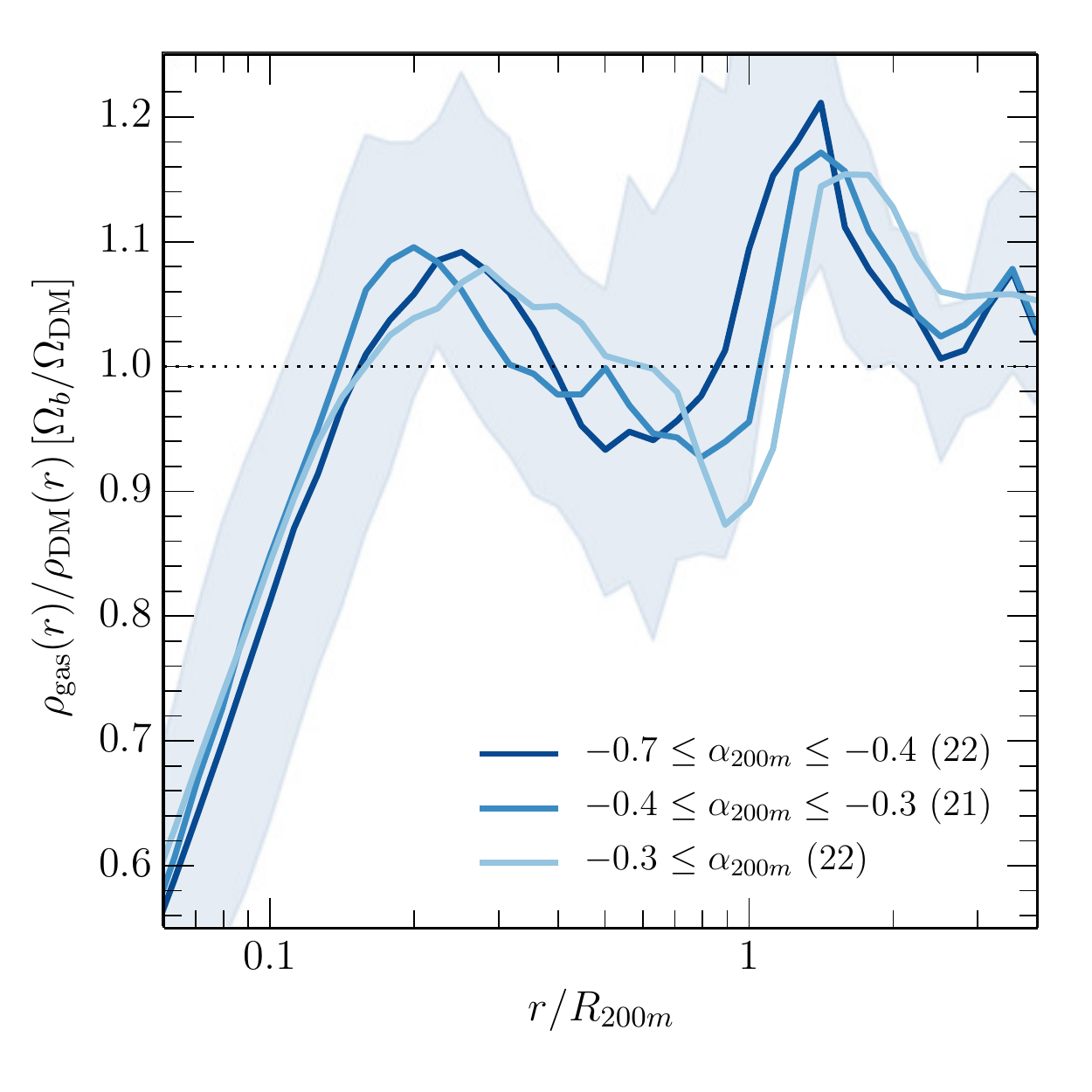}
\caption
{The ratio of mean gas density to mean DM density profile normalized by the cosmic baryon to DM density for different redshifts $z=0, 0.5, 1.0,$ and $1.5$ ({\em left} panel) and for different MAR $\alpha_{200m}$ at $z=0$ ({\em right} panel).  The shaded regions represent 1$\sigma$ scatter in the profile for clusters at $z=0$ ({\em left} panel) and the profile with the most negative $\alpha_{200m}$ ({\em right} panel). The scatter for the other $z$ and $\alpha_{200m}$ bins are similar in size. 
}
\label{fig:rhoratio_r200m}\end{center}
\end{figure*}

Next we study the dependence of the gas profiles on the MAR in the present-day universe at $z=0$.  
We use the MAR proxy $\alpha$ defined in Equation~\ref{eq:alpha} as a probe the MAR of the cluster.  We choose $\Delta_m = 200$ to define $\alpha$ since it scales out the redshift dependence of the MAR, as shown in Section~\ref{sec:redshift_dependence}. 

We first compare our new MAR proxy $\alpha_{200m}$, based on DM infall velocities defined in Equation~\ref{eq:alpha}, with $\Gamma_{200m}$ given by Equation~\ref{eq:gamma}.  
In Figure~\ref{fig:gamma_alpha}, we plot $\alpha_{200m}$ as a function of $\Gamma_{200m}$, with $R_\alpha/R_{200m}= 1.0, 1.25,$ and $1.5$. There is a clear anti-correlation between $\Gamma_{200m}$ and $\alpha_{200m}$ for the above values of $R_\alpha/R_{200m}$, suggesting that the instantaneous MAR proxy $\alpha_{200m}$ is a good alternative probe of $\Gamma_{200m}$.  In particular, we find that setting $R_\alpha=1.25\,R_{200m}$ minimizes the scatter in $\alpha_{200m}-\Gamma_{200m}$, elucidating the clearest dependence of the thermodynamic profiles on the MAR. 

Figure~\ref{fig:vrad_alpha} shows the radial velocity profiles of gas and DM for clusters in three bins of $\alpha_{200m}$, with each bin containing clusters that lie in the top, middle, and bottom third of the distribution in $\alpha_{200m}$.  Not surprisingly, rapidly accreting clusters (with more negative $\alpha_{200m}$) show more negative gas radial velocity in the infall region $0.6\lesssim r/R_{200m}\lesssim 3$. 
The radial gas velocity also becomes progressively less negative, with decreasing size in the infall regions for systems with a less negative $\alpha_{200m}$.  Note that the radial velocity of gas in the infall region is generally less negative than that of DM because the collisional gas slows down at it experiences shocks and ram-pressure of the surrounding ICM. Note, although we treat the DM and gas flows as spherical mass shells moving radially, in reality they are likely to occur anisotropically through mergers and accretions along filaments.  

In Figure~\ref{fig:ent_temp_pres_r200m_z0_vrad} we plot the profiles of entropy, temperature, and pressure for the three different bins of $\alpha_{200m}$. The left panel of the figure shows the entropy profile. The peak of the entropy profile indicates the accretion shock radius $R_{\rm sh}$, which shifts toward the inner regions for higher MAR clusters. This is because more rapidly accreting halos accumulate gas with higher momentum flux, which penetrates deeper into the interior region \citep[e.g.,][]{voit_etal2003, mccourt_etal2013} and shifting the entropy peak and the entropy profile inward. We note that low MAR clusters do not show flatter entropy profile.  This is at odds with the model put forward by \citet{cavaliere_etal2011}, who suggested that the flattening of entropy in cluster outskirts revealed by {\em Suzaku} \citep[e.g.,][]{simionescu_etal2011, walker_etal2013} may be due to weakening of accretion shocks as MAR drops due to the repulsive of dark energy at late cosmic time.  Our simulation shows the opposite. The maximum entropy values of the low MAR clusters are slightly higher than, but still consistent with those of the high MAR clusters.  There is a relatively modest $<20\%$ drop in entropy at $\lesssim R_{200m}$ in the low MAR clusters due to the outward shift of the entropy profile toward larger radii compared to high MAR clusters. This small entropy drop is not enough to explain the observed entropy flattening in {\em Suzaku}. 

Other processes may be responsible for the observed flattening of the entropy profiles in the outskirts of relaxed clusters. For example, the radially dependent ICM inhomogeneities can lead to overestimates in the gas density profiles derived from X-ray observations, causing the observed entropy profile to flatten at the large cluster-centric radii \citep{nagai_lau2011}. Non-thermal pressure due to gas motions induced by gas accretion and mergers \citep[e.g.,][]{vazza_etal2009,nelson_etal2012} as well as plasma effects \citep[e.g., magnetothermal instability,][]{parrish_etal2012} can also bias the gas temperature and entropy profiles low at large cluster-centric radii by keeping some of the gas energy non-thermal. Non-equilibrium electrons can also lower the measured electron temperature inside the accretion shock \citep[e.g.,][]{rudd_nagai2009}, which might partially explain the observed entropy drop in the outskirts \citep{hoshino_etal2010, akamatsu_etal2011}. 

The average gas temperature in more rapidly accreting clusters is lower than the value of the whole cluster sample, especially in their outskirts where gas is actively accreting,  but the impact of MAR on the temperature profiles is smaller within $R_{200m}$ than the entropy profile. At $r=R_{200m}$, the most rapidly accreting clusters have temperature that is about $9\%$ lower than the least accreting systems.  The differences become larger at $r>R_{200m}$, reaching $15\%$ at $r=1.5 R_{200m}$. The gas temperature of high MAR clusters is lower in the outskirts because their accretion shocks are located at smaller radii, and a larger fraction of the outskirt gas is still pre-shocked and has lower temperature. 

Note that our results on the temperature profile and its dependence on MAR are qualitatively different from results based on idealized simulations by \citet{mccourt_etal2013}, who reported that clusters with higher accretion rates have higher temperatures than slowly accreting systems. The discrepancy could be due to their assumption of instantaneous thermalization of accreting gas at the accretion shocks, which could cause the gas temperature to be overestimated. Taking into account the residual kinetic energy from incomplete thermalization in the form of non-thermal pressure could account for this problem.  High MAR clusters are expected to have a higher non-thermal pressure fraction due to the increased level of merger- and accretion-induced gas motions \citep{nelson_etal2014b, shi_komatsu2014, shi_etal2014}. Therefore, the over-predicted gas temperature in the model of \citet{mccourt_etal2013} can be lowered by the correspondingly larger amount of non-thermal pressure present in the high MAR clusters, which could bring their results in better agreement to the results of cosmological simulations. 

On the right panel of Figure~\ref{fig:ent_temp_pres_r200m_z0_vrad} we show the thermal pressure profile and its dependence on MAR. We multiply the pressure profile by the radius cubed to show more clearly the dependence on $\alpha_{200m}$ in the outskirts. Rapidly accreting clusters exhibit lower thermal pressure than slowly accreting systems by $\sim 40\%$, as recently accreted gas is less thermalized and less dense (see Figure~\ref{fig:rho_r200m_z0_vrad}). 

The left panel of Figure~\ref{fig:rho_r200m_z0_vrad} shows the effect of the MAR $\alpha_{200m}$ on gas density. We have multiplied the density profile by the radius squared to show the dependence on $\alpha_{200m}$ more clearly. Similar to the pressure profile, there is a factor of 2 difference in the gas density between the least and most rapidly accreting clusters at $R_{200m}$. The larger inflow in rapidly accreting clusters is responsible for the decrease in density at $ 0.5 \le r/R_{200m}\le 1$. The right panel of Figure~\ref{fig:rho_r200m_z0_vrad} shows the logarithmic gas density slope $\gamma_{\rm gas} \equiv d\log \rho_{\rm gas}/d\log r$. More rapidly accreting clusters have shallower density slopes in gas densities at $0.1\le r/R_{200m}\le 0.3$ and slightly steeper slope at $0.3\le r/R_{200m}\le 1$. The dependence of gas density slope on MAR or the halo formation history is consistent with that of the DM density slope seen in $N$-body simulations \citep{diemer_kravtsov2014,wu_etal2013}. We note, however, that there is a large scatter in the gas density slopes for a given $\alpha_{200m}$ bin.

\subsection{How Does Gas Trace DM in Cluster Outskirts?}
\label{sec:gas_dm}

In this section we examine how well gas density traces DM density in cluster outskirts and its dependence on redshift and MAR. 
The relation between gas and DM densities can be useful for prescribing gas distribution in DM-only simulations, or for inferring
the DM distributions from the observed gas distributions in cluster outskirts. 
Note that in this section, we do not exclude substructures and filaments in either gas or DM components. 

We find that gas traces DM in a uniform manner with redshift when we define clusters with $\Delta_m=200$. 
The left panel of Figure~\ref{fig:rhoratio_r200m} shows the ratio of gas-to-DM density ($\rho_{\rm gas}/\rho_{\rm DM}$) 
as a function of the cluster-centric radius at $z=0.0,0.5,1.0,1.5$. 
Here, the gas and DM densities shown are the mean values in each spherical bin, 
and the ratio is normalized to the cosmic baryon-to-DM ratio ($\Omega_b/\Omega_{\rm DM}$). 
Similar to the thermodynamic profiles discussed in Section~\ref{sec:redshift_dependence}, 
the profile of  $\rho_{\rm gas}/\rho_{\rm DM}$ is universal with $z$ in the radial range $0.3 \le r/R_{200m} \le 2$. 
While the gas density roughly traces the DM density in this radial range,
there is $10\%-20\%$ deviation in the gas-to-DM density ratio from the cosmic value around the accretion shock. 
The value of $\rho_{\rm gas}/\rho_{\rm DM}$ is below the cosmic value in the intermediate region ($0.6 \le r/R_{200m} \le 1$), 
but exceeds the cosmic value in the inner ($0.3 \le r/R_{200m} \le 0.6$) 
and outer ($1.0 \le r/R_{200m} \le 3$) regions of clusters, 
and asymptotically approaches the cosmic value beyond $3 R_{200m}$. 
The deviation from the cosmic value is on average about $10\%$ around the accretion shock, 
but could reach to more than $20\%$ for individual clusters. 
This pattern in the $\rho_{\rm gas}/\rho_{\rm DM}$ profile originates from the difference in the dynamics 
between gas and DM discussed in Section~\ref{sec:redshift_dependence}.  
Shock heating and ram pressure cause gas to lag behind DM during infall, creating
an overdensity of gas around the accretion shock $R_{\rm sh}$, indicated by the peak in 
the $\rho_{\rm gas}/\rho_{\rm DM}$ profile. 
The collisionless DM, on the other hand, penetrates further into the inner region of the cluster, undergoes core passage,  
and accumulates at the first apocenter passage, leading to slightly overdense DM density at $r \sim R_{200m}$. 
This drop in DM densities from $1.0\,R_{200m}$ to $1.6\,R_{200m}$ corresponds to the ``splashback'' radius, where the outermost caustic of the DM is located \citep{diemer_kravtsov2014, adhikari_etal2014}. Our work suggests that the ``splashback'' radius of DM coincides with the accretion shock radius of the gas.  In the right panel of Figure~\ref{fig:rhoratio_r200m}, 
we also show the dependence of MAR in the profiles of the gas-to-DM density ratio for the $z=0$ clusters.  
The profiles follow the same pattern as in the left panel. The peak of the profiles
occurs at smaller radii for clusters with the highest MAR.  

Recently, \cite{patej_loeb2014} proposed an analytical model of gas distribution in galaxy clusters  
that depends on the ratio of the gas density jump to DM density jump at the accretion shock. 
They find that fitting their model to observed gas density profiles infers 
a similar gas density jump to that of DM density around the accretion shock, 
consistent with our findings that the gas density traces DM density 
to within $\lesssim 20\%$ in the cluster outskirts. 
Their model can be further improved by considering the effect
of differential dynamics between the collisionless DM and collisional gas that leads to 
the deviation of the gas-to-DM density ratio from the cosmic mean. 
This will provide an unique approach for constraining the physics of cluster accretion shocks
based on observations of the inner ICM profiles.

\subsection{Location of the Accretion Shock}
\label{sec:rshock}

\begin{figure}
\begin{center}
\includegraphics[scale=0.6]{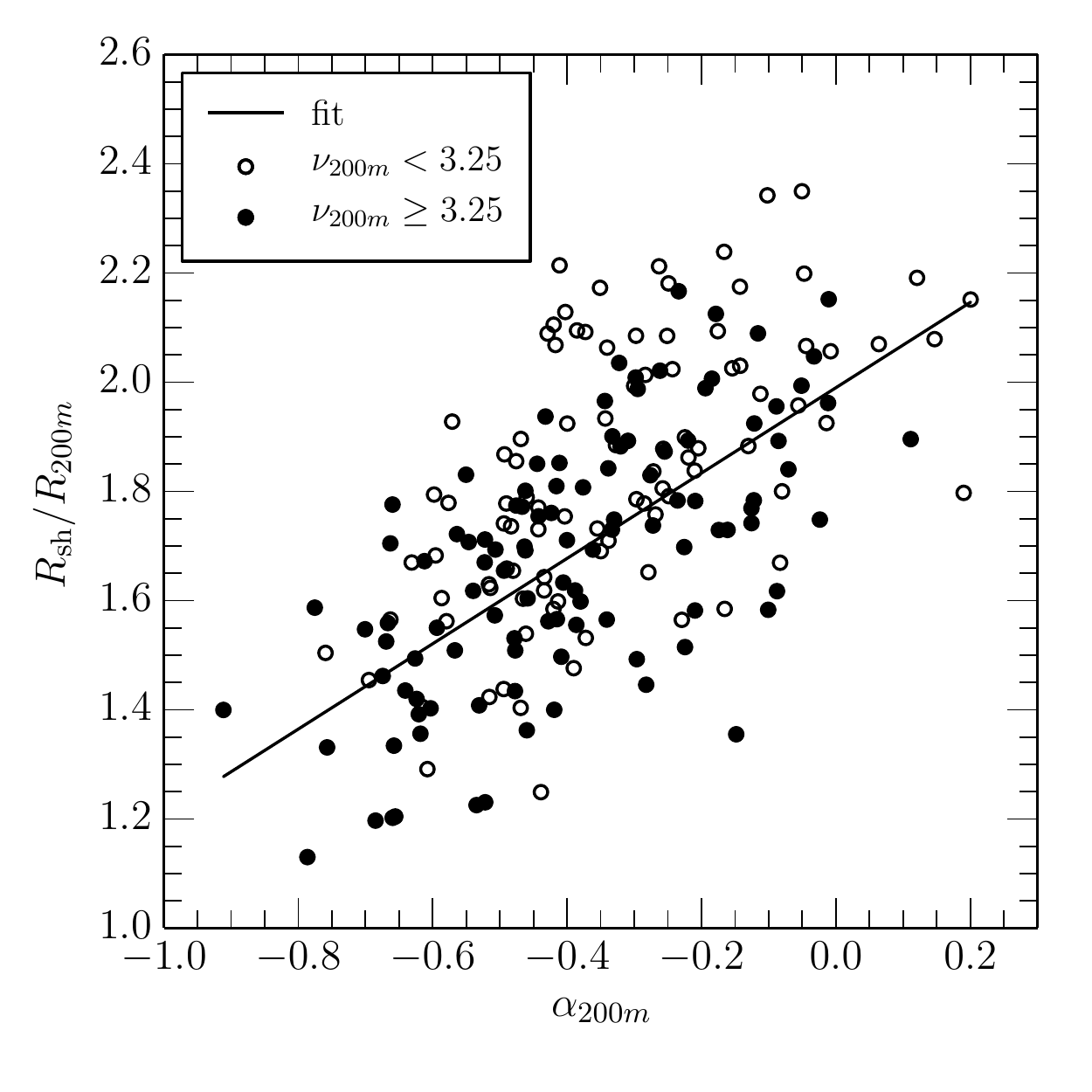}
\caption
{Relation between the shock radius $R_{\rm sh}/R_{200m}$ and the MAR $\alpha_{200m}$ of clusters at $z=0.0,0.5,1.0,1.5$ binned by their peak heights $\nu_{200m}$. The filled and empty points represent clusters with $\nu_{200m}$ greater than or less than $3.25$, respectively. The line represents the best-fit $R_{\rm sh}/R_{200m}$-$\alpha_{200m}$ relation.  }
\label{fig:rshock_alpha}\end{center}
\end{figure}

The dependence of outskirt gas profiles on $\alpha_{200m}$ is similar to that of $z$ for 
$\Delta_c = 200$ (see Figures~\ref{fig:ent_vr_r200c_r200m_z} 
and \ref{fig:rho_temp_pres_r200c_r200m_z}). 
The accretion shock occurs closer to the cluster center for systems with more negative values of $\alpha_{200m}$. 
Similarly, for $\Delta_c = 200$, high-$z$ clusters on average have their accretion shock closer to the cluster center. 
This suggests that the apparent evolution of the profiles for $\Delta_c = 200$ 
originates from the evolution in the {\em physical} MAR, where high-$z$ clusters experience 
more rapid mass accretion than low-$z$ counterparts. Normalizing clusters with respect 
to $\Delta_m = 200$ accounts for the redshift dependence of  effects of the average MAR on the outskirt gas profiles, 
while the residual differences in MAR between clusters at a given redshift contribute to the scatter in the profiles.
We further investigate how the location of the accretion shock depends on redshift, mass, and accretion rate. 
In Figure~\ref{fig:rshock_alpha}, we characterize the relationship between the location of the accretion 
shock  $R_{\rm sh}$ in units of $R_{200m}$ and the MAR proxy $\alpha_{200m}$. 
We divide the cluster sample into a high and low peak height bin, splitting at 
$\nu_{200m} = \delta_c/\sigma(M_{200m},z)=3.25$. The accretion shock radius 
$R_{\rm sh}/R_{200m}$ has a linear correlation that is independent of peak height.   
Halos with different peak height occupy different regions along the relation, 
where low peak height halos tend to have slightly larger $R_{\rm sh}/R_{200m}$ 
and low MAR (more positive $\alpha_{200m}$), although this trend is fairly weak. 

We quantify the best-fit relation between $R_{\rm sh}/R_{200m}$ and $\alpha_{200m}$ by performing linear least square fit:
\begin{equation}
R_{\rm sh}/R_{200m} = A + B \alpha_{200m}
\end{equation}
where $A = 1.990\pm0.030$ and $B= 0.782\pm0.067$. 
Note that this accretion shock radius is considerably larger than $R_{200c}$ ($\approx 0.6 \,R_{200m}$ at $z=0$), 
or the virial radius $R_{\vir}$ ($\approx 0.8\,R_{200m}$ at $z=0$). 
This redshift independent location of the accretion shocks should be useful for modeling how accretion shocks generate non-thermal pressure \citep{shi_komatsu2014},  non-equilibrium electrons \citep[e.g.,][]{avestruz_etal2014b}, and cosmic rays \citep[see][for review]{brunetti_jones2014},
as well as assessing their effects on the hydrostatic mass bias \citep[e.g.,][]{lagana_etal2010}.  Note further that we defined the shock radius 
using the peak of the azimuthally averaged entropy profile, while the actual topology of the accretion shock is quite complicated, which contributes to the 
large scatter in the $R_{\rm sh}/R_{200m}-\alpha_{200m}$ relation. 

\section{Conclusions and Discussion}
\label{sec:conclusions}

In this work we investigated the self-similarity of the diffuse X-ray emitting gas profiles in the outskirts of galaxy clusters using a mass-limited sample of simulated clusters extracted from the {\em Omega500} cosmological hydrodynamic simulation. Our main results are summarized below:

\begin{enumerate}

\item The radial profiles of the diffuse ICM in the outskirts of galaxy clusters at $r\gtrsim R_{200c} \approx 0.6 R_{200m}$ exhibit remarkable self-similarity with redshift when they are normalized with respect to the mean density of the universe, while in the inner regions of clusters they are more self-similar when normalized with respect to the critical density.  This difference in the scaling property of the ICM radial profiles originates from the fact that the outer gas profiles are determined by late time accretion governed by the mean density of the universe, while the inner profiles are determined by the gravitational potential that is set when the universe is still matter-dominated and stays roughly constant afterward. 
  
\item The diffuse ICM profiles in cluster outskirts depend on the mass accretion rate (MAR) of the cluster.  To quantify the MAR, we proposed a new indicator $\alpha$ which is defined as the ratio of the local infall dark matter (DM) velocity in the halo outskirt to the virial velocity of the halo. Using this new indicator, we find that the pressure and temperature profiles of low MAR clusters are systematically higher than those of high MAR clusters, because a significant fraction of kinetic energy associated with accreting materials has not yet been thermalized in rapidly accreting clusters. Specifically, our results suggest that the ICM pressure profile is not ``universal'' at large radii when scaled to ${R}_{200c}$. While the pressure profile in cluster outskirts exhibits a more universal evolution when scaled to $R_{200m}$, the profile at large radii is generally sensitive to MAR.  Therefore, any use of the ``universal'' pressure profile in extrapolating thermal SZ measurements from the outskirts to $R_{500c}$ will likely be biased.  This is especially important when the beam size of the instrument is too large to resolve $R_{500c}$ for high-$z$ clusters, e.g., as in the case of {\em Planck}. 
Our work suggests that the effects of MAR must be taken into account when interpreting SZ observations of cluster outskirts, including the recent {\em Planck}'s stacked SZ measurements which detected thermal pressure profiles around massive clusters out to $3 \times R_{500c} \approx 1.2\,R_{200m}$ \citep{planck_interV2013}. 

\item Gas does not trace DM perfectly in the infall regions of galaxy clusters, because the collisional gas accretes slower than the collisionless DM.  This causes the gas-to-DM density ratio to deviate from the cosmic mean value by about $10\%$ near the accretion shocks, and steepens the gas density profile relative to the DM profile at large cluster-centric radii.  Recent ultra-deep ($\gtrsim 2$~Ms) {\em Chandra} observation of Abell 133 and Abell 1795 (Vikhlinin~et~al., in~prep.) may be able to detect the steepening in X-ray emissions in the diffuse ICM component, after properly removing point sources, clumps, and filaments.

\item The accretion shock radius $R_{\rm sh}$ is on average located at the fixed fraction of $R_{200m}$ ($R_{\rm sh}/R_{200m} \approx 1.6$) of galaxy clusters independent of redshift ($0\le z \le 1.5$). However, there is also a large scatter in the accretion shock radius ($R_{\rm sh}/R_{200m}  \approx1.0 - 2.4$), depending on the MAR of clusters. Higher MAR clusters have smaller accretion shock radius.  These results can be useful for modeling physical processes (such as generation of turbulence and cosmic-rays) related to accretion and shock heating at outer boundaries of galaxy clusters. 

\item Our results suggest that the critical density is still preferred in defining cluster mass and radius, for calibrations of observable-mass relations (e.g., $M-Y_X$ and $M-Y_{SZ}$) based on the current generation of X-ray and SZ profile measurements, which mostly probe gas out to $r \lesssim R_{500c}$.  Since the outer profiles are more self-similar when they are normalized with respect to the mean mass density, the exploitation of cluster outskirts for cosmology requires some care.  For example, using the critical density in normalizing the outer ICM profiles can introduce redshift-dependent systematic biases in cluster scaling relations.  Furthermore, scaling relations of cluster outskirts are expected to show larger scatter due to variations in MAR.  Detailed understanding of physical processes and observational biases will be critical for interpreting data from the next-generation of X-ray and SZ missions, such as {\em SMART-X}\footnote{\url{http://smart-x.cfa.harvard.edu/}} and {\em Athena}+.\footnote{\url{http://athena2.irap.omp.eu/}}

\end{enumerate}

\acknowledgements We thank Benedikt Diemer, Oleg Gnedin, Eiichiro Komatsu, Andrey Kravtsov, Avi Loeb, Xun Shi, Andrew Wetzel, 
and the anonymous referee for useful discussion and/or comments on the manuscript. This work was supported in 
part by NSF grants AST-1412768 \& 1009811, NASA ATP grant NNX11AE07G, NASA Chandra grants GO213004B 
and TM4-15007X, the Research Corporation, and by the facilities and staff of 
the Yale University Faculty of Arts and Sciences High Performance Computing Center. 
CA acknowledges support from the NSF Graduate Student Research Fellowship 
and Alan D. Bromley Fellowship from Yale University. 

\vspace{-5pt}

\bibliography{ms}

\end{document}